\documentclass[aps,prx,reprint,preprintnumbers,superscriptaddress,nofootinbib,longbibliography,floatfix]{revtex4-2}
\pdfoutput=1
\usepackage{rotating}
\usepackage{array}
\usepackage{amsmath}
\usepackage[normalem]{ulem}
\usepackage{slashed}
\usepackage{booktabs}
\usepackage[pdftex,table]{xcolor}
\usepackage{units}
\usepackage{xfrac}
\usepackage{mathtools}
\usepackage{empheq}
\usepackage[]{units}
\usepackage{multirow}
\usepackage{amssymb}
\usepackage{url}
\usepackage{comment}
\usepackage{physics}
\usepackage{color,soul}
\usepackage{bbm}
\usepackage[caption=false]{subfig}
\usepackage{graphicx}

\usepackage{hyperref}
\hypersetup{
  colorlinks=true,
  citecolor=blue,
  linkcolor=blue,
  urlcolor=blue
}

\newcommand{\pt}{\mathrm{p_{T}}}

\begin{document}

\title{Full Phase Space Resonant Anomaly Detection}

\author{Erik Buhmann}
% \email{erik.buhmann@uni-hamburg.de}
\affiliation{Institute for Experimental Physics, Universität Hamburg, Luruper Chaussee 149, 22761 Hamburg, Germany}

\author{Cedric Ewen}
\email{cedric.ewen@studium.uni-hamburg.de}
\affiliation{Institute for Experimental Physics, Universität Hamburg, Luruper Chaussee 149, 22761 Hamburg, Germany}

\author{Gregor Kasieczka}
%\email{gregor.kasieczka@puni-hamburg.de}
\affiliation{Institute for Experimental Physics, Universität Hamburg, Luruper Chaussee 149, 22761 Hamburg, Germany}

\author{Vinicius Mikuni}
\email{vmikuni@lbl.gov}
\affiliation{National Energy Research Scientific Computing Center, Berkeley Lab, Berkeley, CA 94720, USA}

\author{Benjamin Nachman}
%\email{bpnachman@lbl.gov}
\affiliation{Physics Division, Lawrence Berkeley National Laboratory, Berkeley, CA 94720, USA}
\affiliation{Berkeley Institute for Data Science, University of California, Berkeley, CA 94720, USA}

\author{David Shih}
%\email{dshih@physics.rutgers.edu}
\affiliation{New High Energy Theory Center, Rutgers University \\
  Piscataway, New Jersey 08854-8019, USA}

\begin{abstract}
   Physics beyond the Standard Model that is resonant in one or more dimensions has been a longstanding focus of countless searches at colliders and beyond. Recently, many new strategies for \textit{resonant anomaly detection} have been developed, where sideband information can be used in conjunction with modern machine learning, in order to generate synthetic datasets representing the Standard Model background.  
   Until now, this approach was only able to accommodate a relatively small number of dimensions, limiting the breadth of the search sensitivity.  Using recent innovations in point cloud generative models, we show that this strategy can also be applied to the full phase space, using all relevant particles for the anomaly detection.  As a proof of principle, we show that the signal from the R\&D dataset from the LHC Olympics is findable with this method, opening up the door to future studies that explore the interplay between depth and breadth in the representation of the data for anomaly detection.
\end{abstract}

\maketitle

\section{Introduction}
\label{sec:intro}

In the pursuit of physics beyond the Standard Model (BSM), it is apparent that a parallel path is required to the traditional signal-specific paradigm.  Many physics models have too many parameters to scan efficiently and often specific assumptions are still required to interpret the scanned parameter space~\cite{ATLAS:2015wrn,CMS:2016lcl,Berger:2008cq,Cahill-Rowley:2012ydr,Ambrogi:2017lov,AbdusSalam:2011fc}. Additionally, it may be that new phenomena are not described by known physics models.  While most BSM searches posit a specific model or models in the design and execution of an analysis, \textit{anomaly detection} is an alternative strategy that tries to be agnostic to any particular signal model.

Modern machine learning has accelerated the development and deployment of anomaly detection methods to particle physics~\cite{Kasieczka:2021xcg,Aarrestad:2021oeb,Karagiorgi:2021ngt}.  One of the biggest challenges with any anomaly detection method is the need to balance signal sensitivity with background specificity, i.e. it is necessary to be able estimate the background in any signal-sensitive region.  This is one reason that resonant anomaly detection has attracted significant attention~\cite{Collins:2018epr,Collins:2019jip,Nachman:2020lpy,Andreassen:2020nkr,Stein:2020rou,Amram:2020ykb,Hallin:2021wme,Collins:2021nxn,1815227,Kasieczka:2021tew,Hallin:2022eoq,Chen:2022suv,Kamenik:2022qxs,Sengupta:2023xqy,Raine:2022hht,Golling:2023yjq, bickendorf2023combining, finke2023roots}: sideband regions can naturally be used to estimate backgrounds.

While a number of anomaly detection methods have been applied to resonances, the most sensitive approaches use weakly-supervised learning techniques that are tailored for the resonant case~\cite{Amram:2020ykb,Kasieczka:2021xcg,Collins:2021nxn}.  Of these, nearly all use only a handful of features to avoid sculpting the resonant feature and/or to cope with previous computational limitations.
%~\cite{Collins:2018epr,Collins:2019jip,Nachman:2020lpy,Andreassen:2020nkr,Stein:2020rou,Hallin:2021wme,1815227,Hallin:2022eoq,Chen:2022suv,Kamenik:2022qxs,Sengupta:2023xqy,Raine:2022hht,Golling:2023yjq, bickendorf2023combining, finke2023roots}.  
Feature selection implicitly introduces a prior over theoretical models, as only certain signals would be observable with the chosen features.  While general symmetry considerations can provide the basis for a model agnostic feature reduction, the number of relevant features required to maintain broad sensitivity is still large~\cite{Dillon:2022tmm,finke2023roots}.

We propose a strategy to perform weakly supervised, resonant anomaly detection using a maximal feature set --- all of the relevant particles in an event.  Our approach is based on the \textsc{Cathode} method introduced in~\cite{Hallin:2021wme}: first, conditional generative models are trained on sideband regions and interpolated into the signal region in order to create a synthetic sample of events following the background distribution; second, a classifier is trained to distinguish between background and data, and the classifier output serves as the anomaly score. In~\cite{Hallin:2021wme}, the method was demonstrated using an ordinary normalizing flow trained on a small set of high level features (masses and $n$-subjettinesses). Here, we extend \textsc{Cathode} using recent innovations in point cloud generative models in order to generate the full set of low-level features -- the four momenta of all of the jet constituents.

We will compare two different state-of-the-art approaches to particle cloud generation: a score-based diffusion model~\cite{score_denoising,scoremodels} parametrized by transformer layers~\cite{vaswani2023attention} for permutation equivariance~\cite{mikuni:point_clouds,Mikuni:2023tok}; and a continuous normalizing flow (CNF)~\cite{DBLP:journals/corr/abs-1806-07366} trained with flow-matching~\cite{lipman2023flowmatching} using EPiC layers~\cite{Buhmann:2023pmh} for permutation equivariance. 

We will see that both generative models have comparable performance on the anomaly detection task. They are able to use all of the particles inside jets to identify the anomaly in the R\&D dataset of the LHC Olympics data challenge~\cite{Kasieczka:2021xcg,LHCOlympics}. Currently, our approach is limited in the sense that the sensitivity to the anomaly is only present when the BSM cross section is quite large. However, we expect this sensitivity to improve with future innovations with the generative modeling and/or the classifier. We expect that searching over so many dimensions enhances the breadth of sensitivity at the cost of depth of sensitivity to particular physics models.  Future studies will explore the details of the breadth/depth tradeoff. 

This paper is organized as follows.  Section~\ref{sec:dataset} briefly introduces the dataset. Our machine learning methods are described in Sec.~\ref{sec:method}.  The numerical results are presented in Sec.~\ref{sec:results} and the paper ends with conclusions and outlook in Sec.~\ref{sec:conclusions}.

\section{Dataset}
\label{sec:dataset}

To illustrate our approach, we use the R\&D dataset from the LHC Olympics data challenge~\cite{LHCOlympics,Kasieczka:2021xcg} which is briefly described in the following.  The background is modeled as inclusive dijets and the signal is resonant boson production $A \to B (\to q q') C(\to q q') $ with masses $m_{A}, m_B, m_C$ = 3.5, 0.5, 0.1~TeV, respectively.  The events are generated with \textsc{Pythia8}~\cite{Pythia2} and \textsc{Delphes3.4.1}~\cite{delphes1, delphes2, delpes3} provides a detector simulation. Jets are defined using the anti-$k_T$ jet clustering algorithm~\cite{anti-kt} as implemented in \textsc{FastJet}~\cite{fastjetManual} with $R=1$.  The focus is on events defined by two high energy jets capturing the decay products of the $B$ and $C$ particles.  The pair of jets has a large dijet mass (corresponding to $m_A$) and non-trivial substructure (corresponding to the two-prong decay).  Consequently, the leading jet is required to have $p_T > 1.2$~TeV.

Each event is represented by up to 700 particles with a three-momentum per particle; after selection of the two most energetic jets, the maximum particle multiplicity observed is reduced to 279 per jet. For many previous studies, this high-dimensional information was compressed into a small set of features, such as the masses of the leading jets and the $n$-subjettiness observables $\tau_1$ and $\tau_2$~\cite{nsubjetiness1,nsubjetiness2} of the two jets.  Instead, we consider all of the particles within the leading two jets. For applications at the LHC, we envision a similar approach using inputs based on reconstructed objects such as Unified Flow Objects~\cite{ATLAS:2020gwe} in ATLAS or particle flow inputs~\cite{CMS:2017yfk} in CMS.

\section{Methods}
\label{sec:method}

We introduce two approaches for generating the dijet system. Both approaches are based on the same structure: one model generates overall jet properties and a second model generates the jet constituents conditioned on the overall features. 
For the first model, each jet is represented by its transverse momentum $\pt$, pseudo-rapidity $\eta$, azimuthal angle $\phi$, jet mass $m$, and constituent particle multiplicity $N$. Jet constituents (particles) are generated in the relative set of coordinates defined by $(\pt_\text{rel}, \Delta\eta,\Delta\phi)$, where $\pt_\text{rel} = \pt_\text{part.}/\pt_\text{jet},\Delta\eta = \eta_\text{part.} - \eta_\text{jet}$ and $\Delta\phi = \phi_\text{part.} - \phi_\text{jet}$.

Similarly to previous studies, we define the signal region of interest for events with $3300~\text{GeV} < m_{jj} < 3700~\text{GeV}$. The events used for the training in the sidebands are restricted between the interval $2300~\text{GeV} < m_{jj} < 3300~\text{GeV}$ and $3700~\text{GeV} < m_{jj} < 5000~\text{GeV}$. The interval range for the sideband region is defined such that enough training data is available for the generative model while avoiding boundary effects at the extremes of the dijet mass distribution. Moreover, 60\% of the events in the sideband regions are used for training and testing, while 40\% are used for the validation and further refinement of the generative model. In order to get the $m_{jj}$ values in the SR, we perform a kernel density estimate (KDE) fit to the training data. The KDE was implemented with a gaussian kernel and a bandwidth of $0.001$ using
the \textsc{Scikit-learn}~\cite{scikit-learn} library.

\subsection{Diffusion Generative Models}
\begin{figure*}[ht]
\centering
    \includegraphics[width=0.9\textwidth]{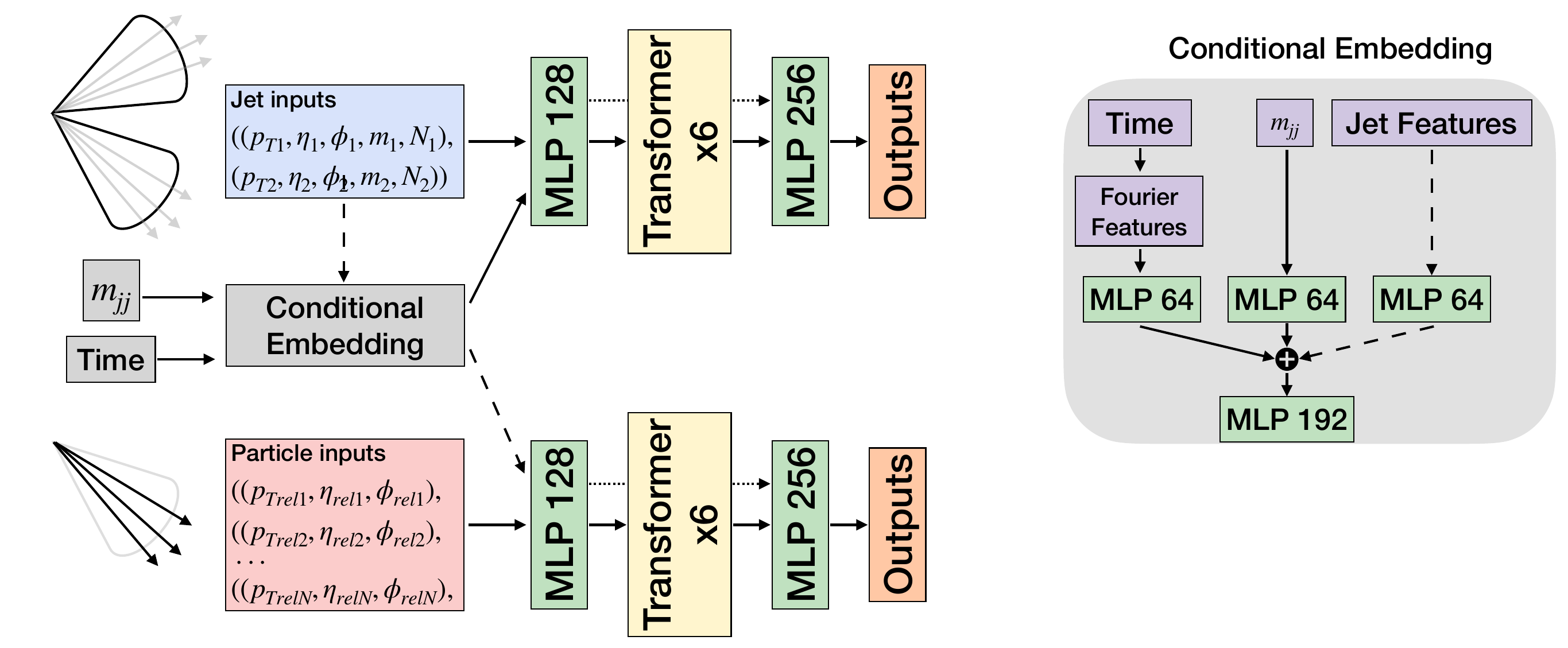}
\caption{Network architecture of the diffusion generative model. Dijet kinematic information is first generated with a dedicated diffusion model. For each jet created, particles are generated through a second diffusion process conditioned on the jet kinematic information. A conditional embedding is used to process the conditional inputs used for both generative models consisting of the invariant mass of the dijet system ($m_{jj}$), the time information, and the jet kinematic information in the case of the particle generation model.}
\label{fig:diagram}
\end{figure*}

Our first generative approach is based on the diffusion models introduced in Ref.~\cite{mikuni:point_clouds}.
The neural network architecture used for both the dijet system generation and particle generation are based on a combination of \textsc{DeepSets}~\cite{DBLP:journals/corr/ZaheerKRPSS17} with attention layers~\cite{DBLP:journals/corr/VaswaniSPUJGKP17} such that permutation equivariance for the network outputs is guaranteed both at the jet- and particle-generation level. Both diffusion models improve the sensitivity to the time information  by feeding random Fourier features \cite{tancik2020fourier} through two fully connected layers. This conditional information is then combined with values of $m_{jj}$ after a log-transformation. In the case of the particle generation model, the additional jet kinematic information is passed through a fully connected layer before being concatenated with the other conditional inputs. A diagram showing the interplay between the generative models is shown in Fig.~\ref{fig:diagram}. Hyperparameter choices are listed in Tab.~\ref{tab:hyperparameters}.

To generate events, we use the trained diffusion models with the DDIM~\cite{DBLP:journals/corr/abs-2010-02502} sampler with 512 time-intervals and additional second-order correction~\cite{karras2022elucidating} to improve precision. A simple pre-processing is applied to all inputs to the generative models to ensure all distributions, for particle and jet kinematics, are transformed to have zero mean and unit variance. The training is carried out using the Perlmutter supercomputer interfaced with the Horovod package~\cite{sergeev2018horovod} for distributed training. 16 NVIDIA A100 GPUs are used simultaneously during training, while a single GPU is used for evaluation. All models are trained for up to 300 epochs with a cosine learning rate schedule~\cite{DBLP:journals/corr/LoshchilovH16a} with initial learning rate of $1.6\times10^{-3}$. If the loss function does not decrease for 30 consecutive epochs, evaluated in a separate testing set representing 10\% to the sample size, the training is stopped. The implementation of all models is carried out using \textsc{Keras} backend~\cite{keras} with \textsc{TensorFlow}~\cite{tensorflow}. 

\subsection{Flow Matching Generative Models}
The second approach is based on the optimal transport flow matching training objective for training continuous normalizing flows (CNFs)\cite{DBLP:journals/corr/abs-1806-07366} introduced in Ref.~\cite{lipman2023flowmatching}. For the jet feature model, the architecture is a fully connected network. 
The particle feature model is based on the EPiC-FM model introduced in Ref.~\cite{Buhmann:2023zgc} and combines the flow matching objective with fast and linearly scalable equivariant point cloud (EPiC) layers~\cite{Buhmann:2023pmh}. Before feeding the point cloud into the EPiC layers, the data is projected into a higher dimensional representation. Because EPiC layers require additional global features as inputs, these are calculated with an average and summation pooling. 

\begin{figure*}[ht]
\centering
    \includegraphics[width=0.9\textwidth]{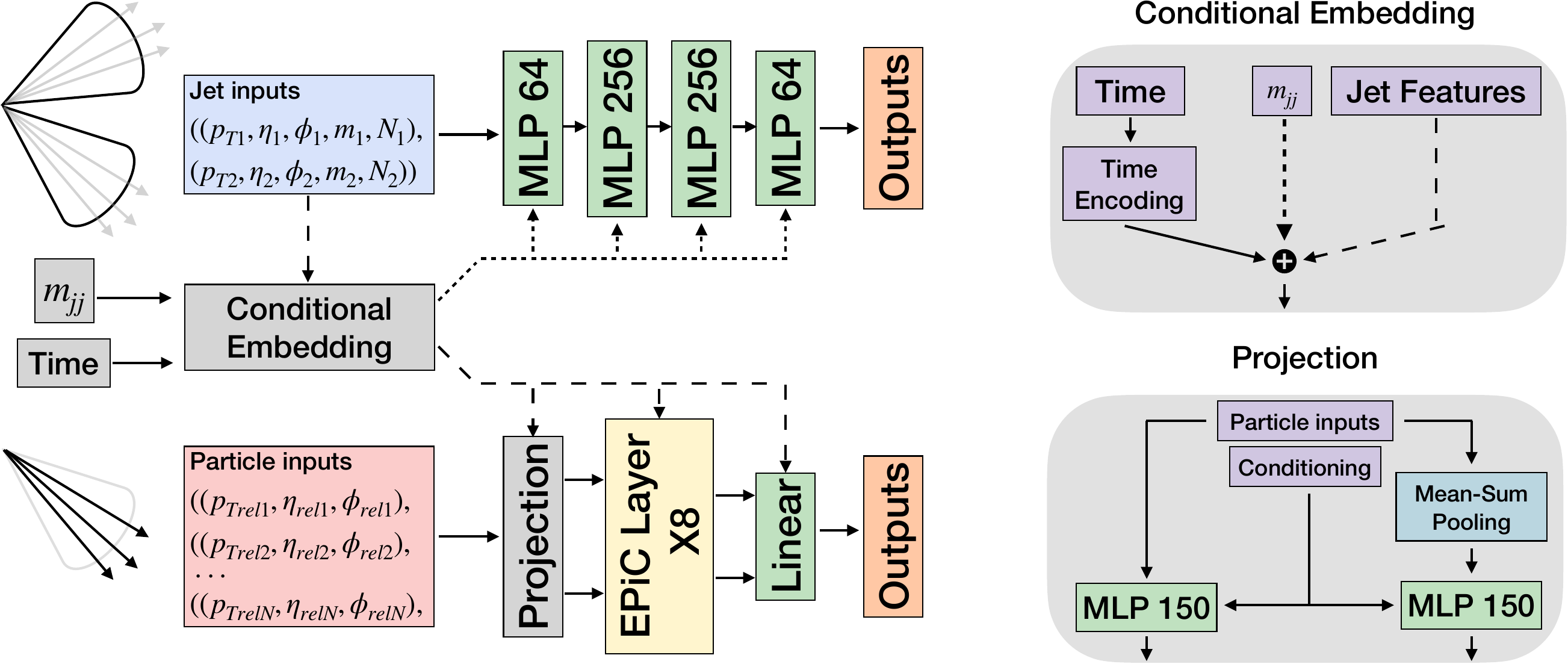}
\caption{Network architecture of the flow matching generative models. The upper architecture first generates the kinematic dijet information, the lower architecture generates the particles of each jet based on the jet kinematic information. The jet feature model is conditioned on time $t$ and dijet mass $m_{jj}$, while the particle feature model is explicitly conditioned on the jet features $\pt$, $\eta$, $\phi$, and $m$. $N$ is used to determine the number of particles per jet.}
\label{fig:diagram_fm}
\end{figure*}

Both models incorporate an embedded time vector and interpolation into the signal region is possible due to the conditioning of the jet feature model on $m_{jj}$. The conditioning for the particle feature model consists of the embedded time and the jet features $\pt$, $\eta$, $\phi$, and $m$. In both models the conditioning is added to all multilayer perceptrons (MLPs).
A visualization of the model architecture can be found in Fig.~\ref{fig:diagram_fm}, and for an overview of all hyperparameter choices we refer to Tab.~\ref{tab:hyperparameters}.

In the generation process, the second order midpoint ordinary differential equation solver with 250 steps (500 function evaluations) from the \textsc{Torchdyn}~\cite{torchdyn} library is used. As for the diffusion models, a simple pre-processing ensures all distributions to have zero mean and unit variance.
The models are implemented with \textsc{PyTorch}~\cite{pytorch} using the \textsc{PyTorch Lightning}~\cite{Falcon_PyTorch_Lightning_2019} framework. 

\subsection{Classifier}

\begin{figure*}[ht]
\centering
    \includegraphics[width=0.8\textwidth]{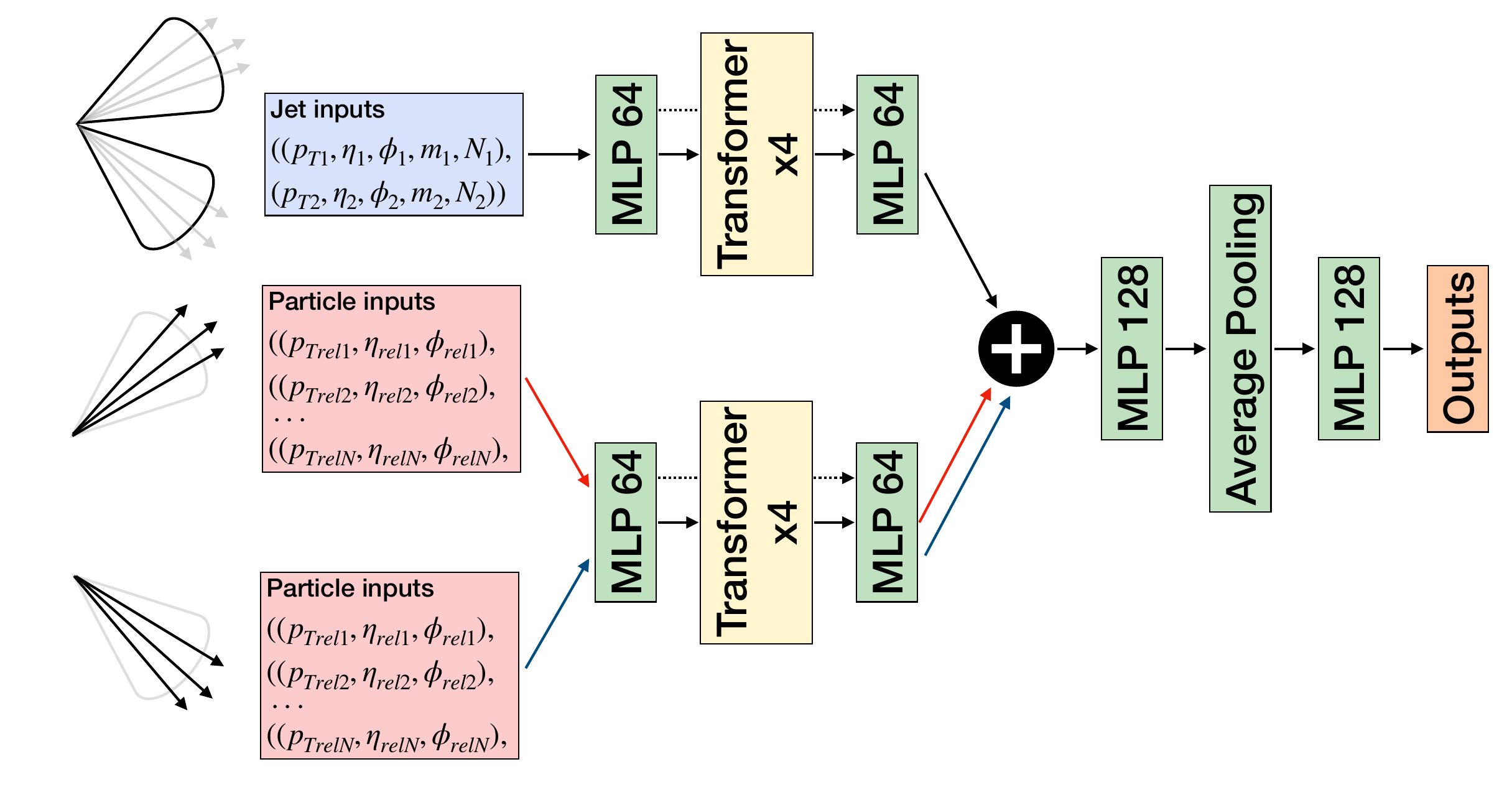}
\caption{Network architecture of the classifier model. Dijet kinematic information is first processed with a transformer model. For each jet, particles are processed through a second transformer model shared between the two jets. The outputs of the transformer model are then combined through a concatenation operation.}
\label{fig:classifier}
\end{figure*}

To obtain the final anomaly score, we employ classifiers trained to distinguish the generated, synthetic background samples from the data - this is the weakly-supervised anomaly detection approach proposed in Ref.~\cite{Hallin:2021wme} following the CWoLa framework of
Refs.~\cite{Metodiev:2017vrx,Collins:2018epr,Collins:2019jip}. In this work, we employ a transformer-based classifier architecture with nearly the same structure as the transformer used in the diffusion model. It first takes the dijet system consisting of the jet kinematic information and uses \textsc{DeepSets} combined with attention layers to map the jets to a common latent embedding where the clustered particles are also represented after applying a similar set of \textsc{DeepSets} layers with added attention. The attention layers used to process the particle information consider in the attention matrix only particles belonging to the same jet, with same set of functions reused between the two jets to ensure the entire classifier is permutation invariant with respect to the order of the input jets and the order of the input particles belonging to each jet. In the same latent embedding space, the dijet system and particles are then concatenated in the particle dimension, resulting in a hybrid graph with $N_1+N_2+2$ ``particles'',  represented by the two jets containing $N_1$ and $N_2$ particles, respectively. A fully connected layer is then added after the concatenation, operating only at the feature level, followed by a dropout operation~\cite{JMLR:v15:srivastava14a} and \textsc{LeakyRelu} activation function. An average pooling operation is then used in the particle dimension to combine the information of all the objects. A second fully connected layer with dropout and \textsc{LeakyRelu} activation are used before the output layer with a fully connected layer of single output and \textsc{sigmoid} activation function. The classifier shares the same learning rate schedule and early stopping criteria as the diffusion models. A visual representation of the classifier architecture is shown in Fig.~\ref{fig:classifier}.

\section{Results}
\label{sec:results}

\begin{figure*}[ht]
\centering
    \includegraphics[width=0.23\textwidth]{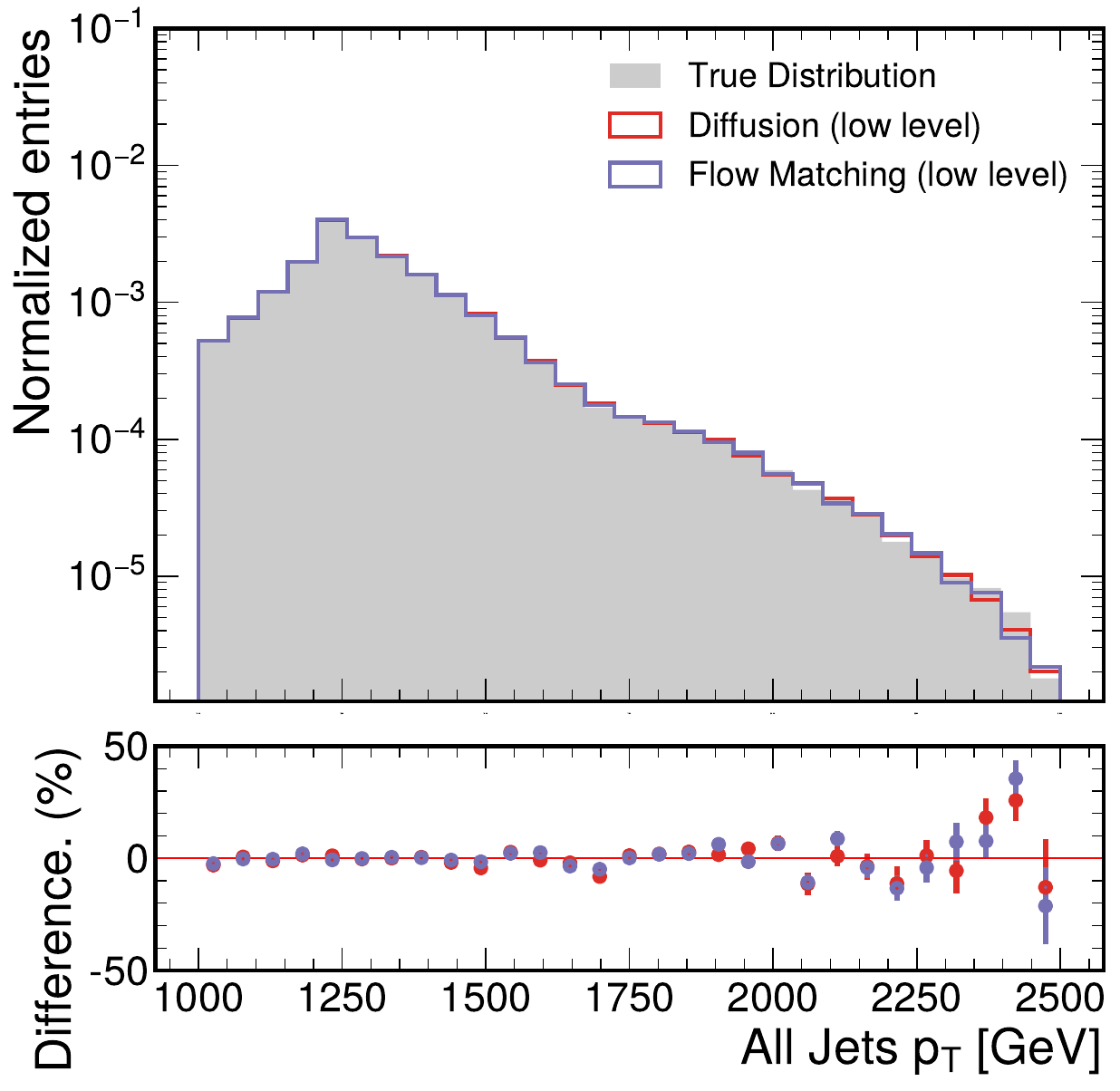}
    \includegraphics[width=0.23\textwidth]{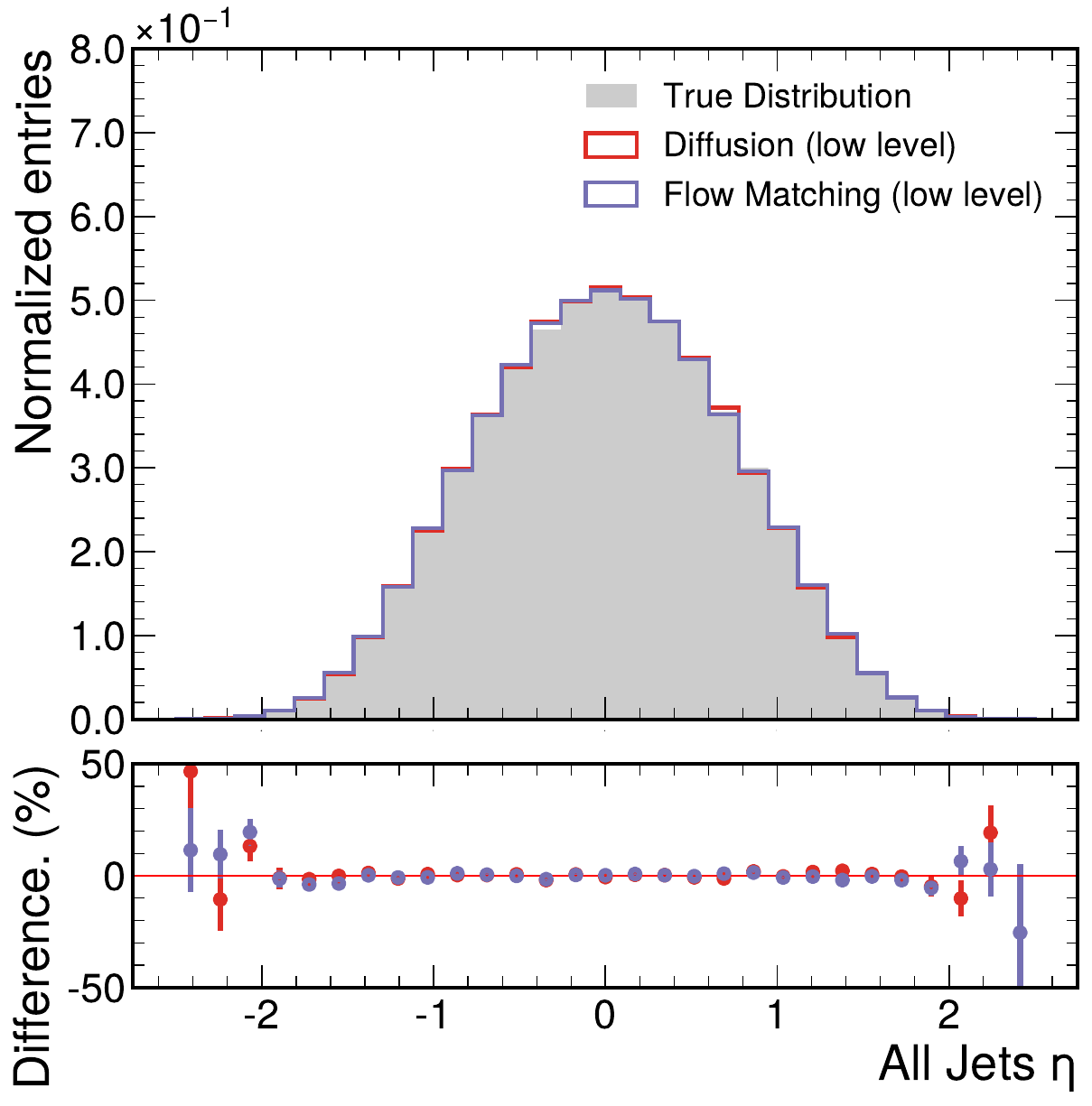}
    \includegraphics[width=0.23\textwidth]{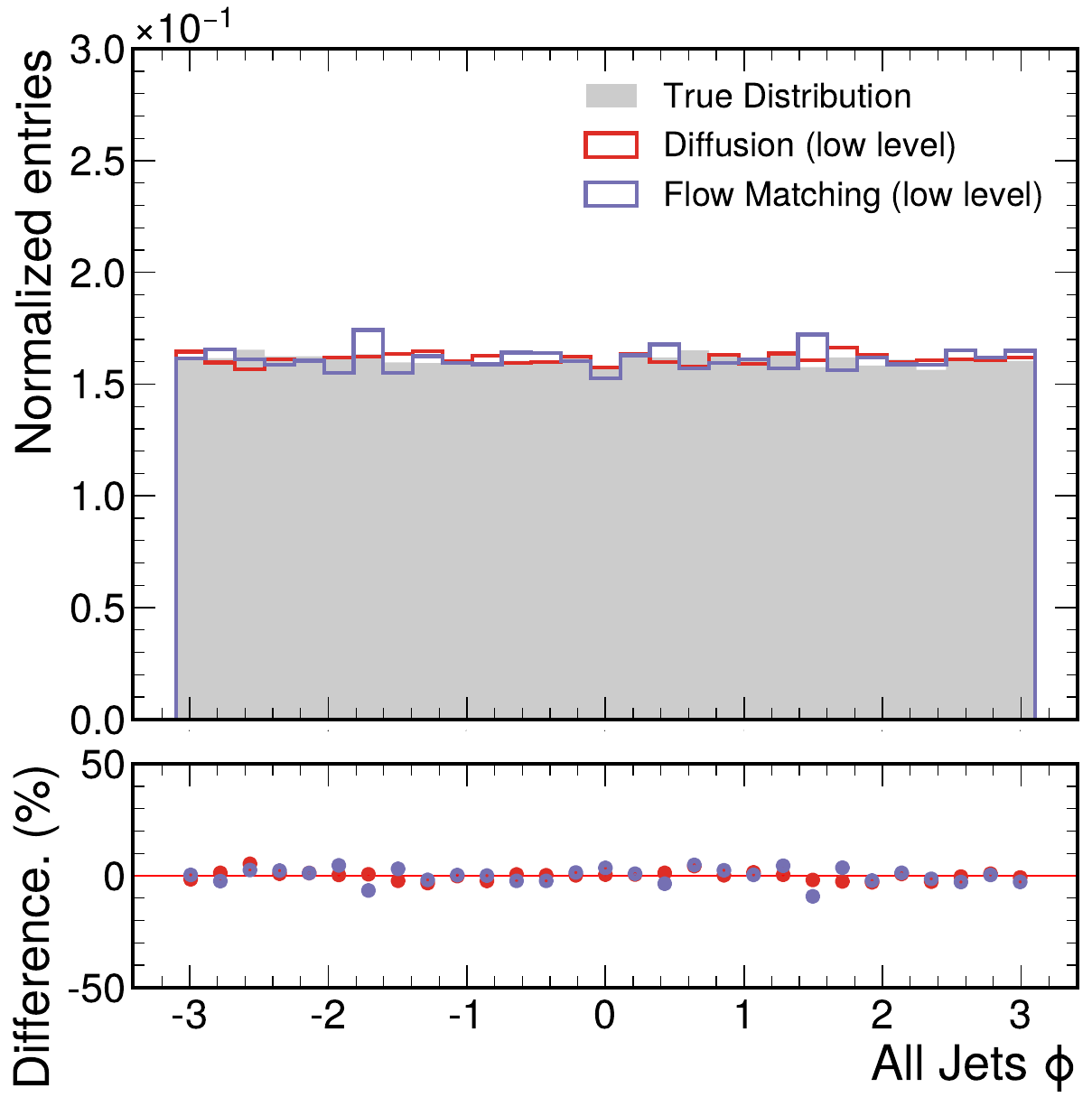}
    \includegraphics[width=0.23\textwidth]{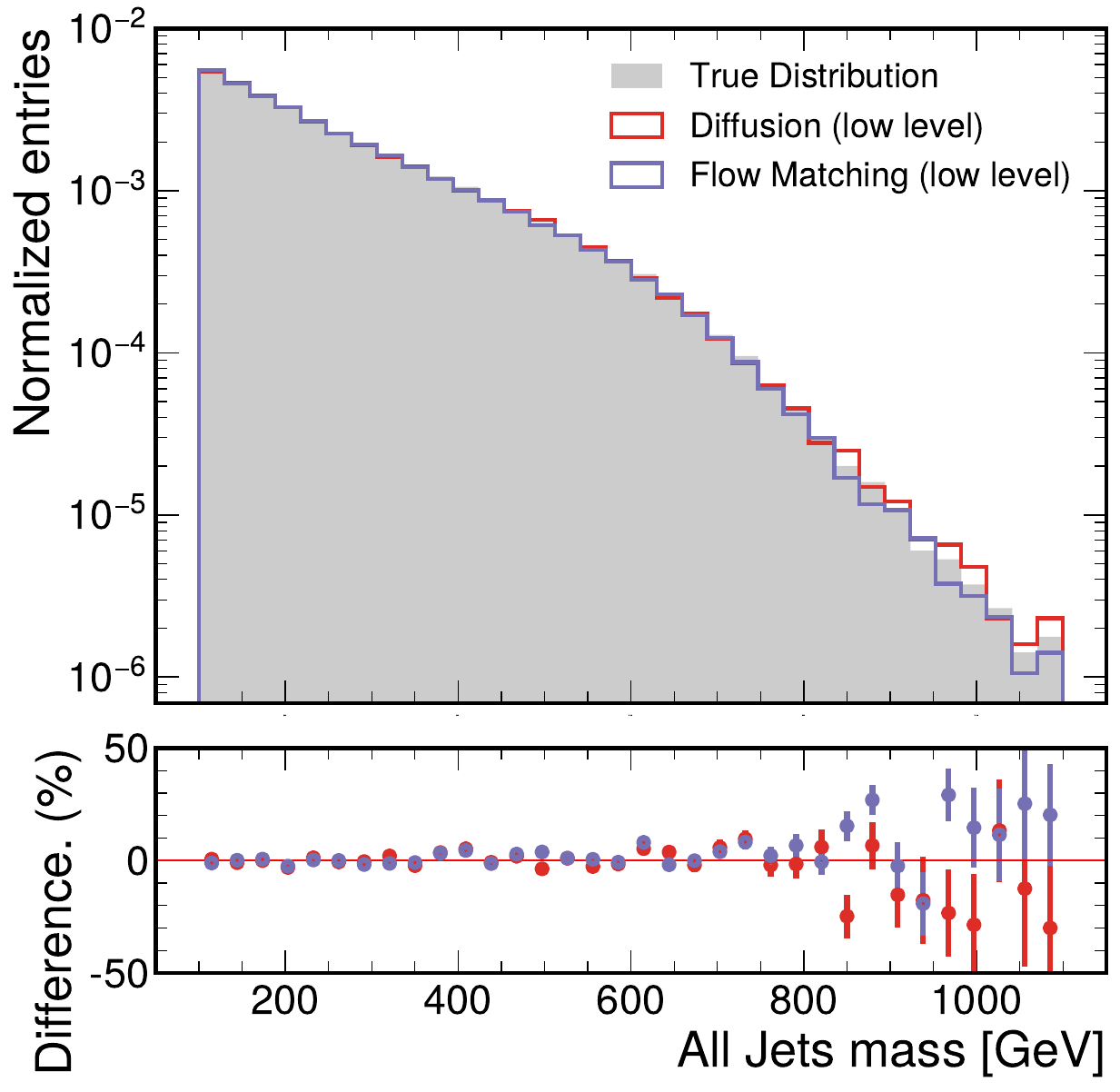}
    \includegraphics[width=0.23\textwidth]{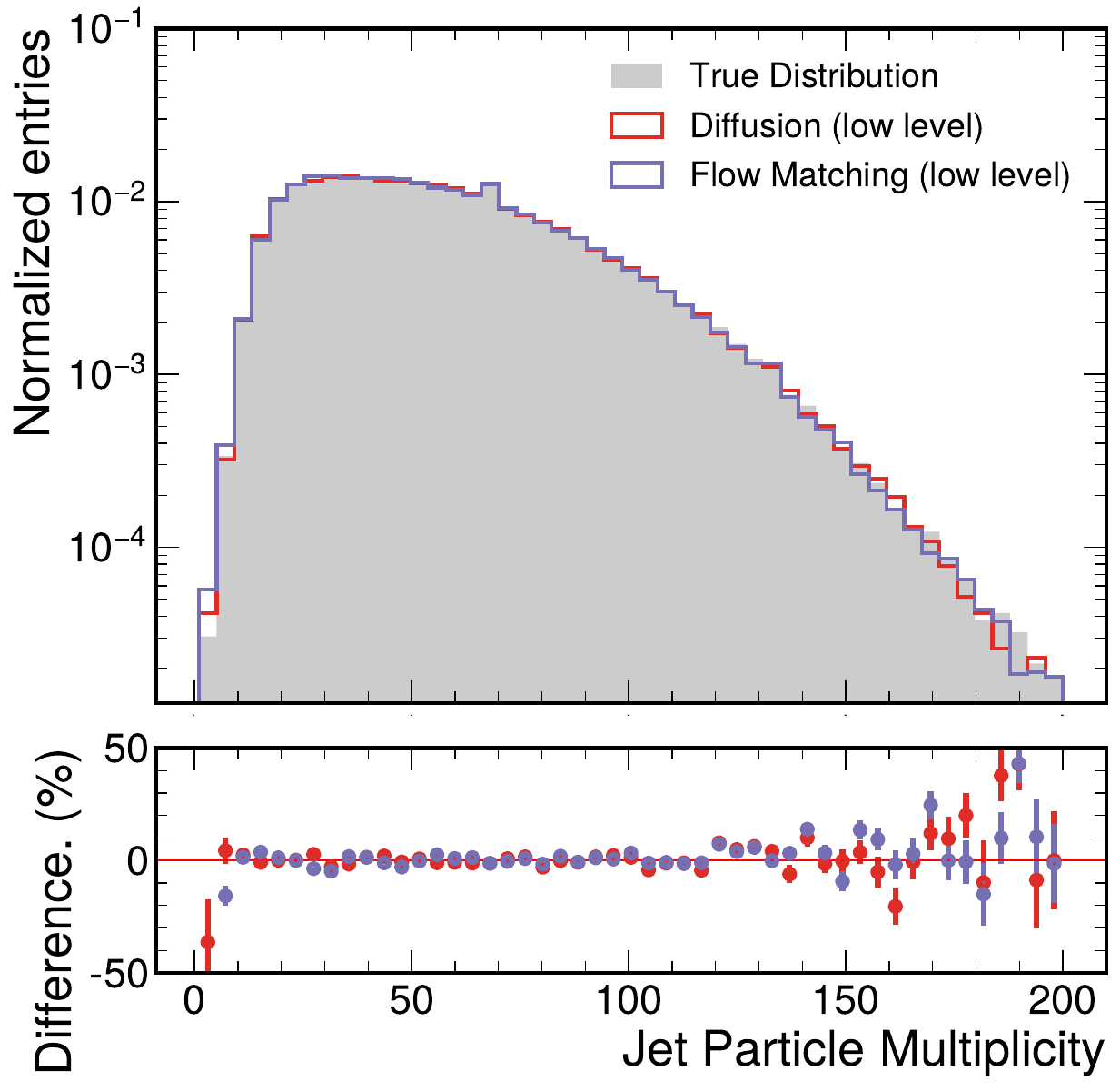}
    \includegraphics[width=0.23\textwidth]{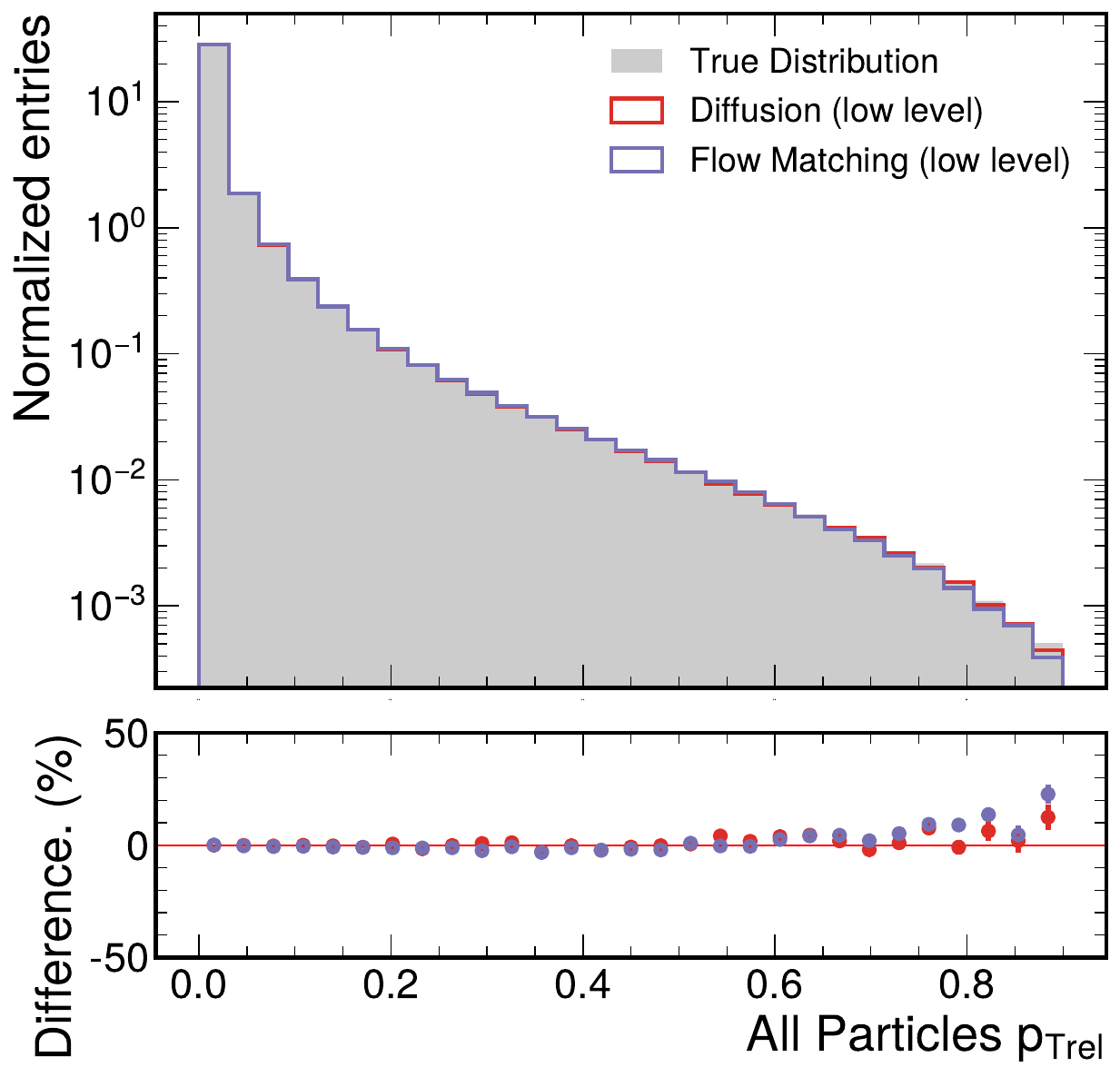}
    \includegraphics[width=0.23\textwidth]{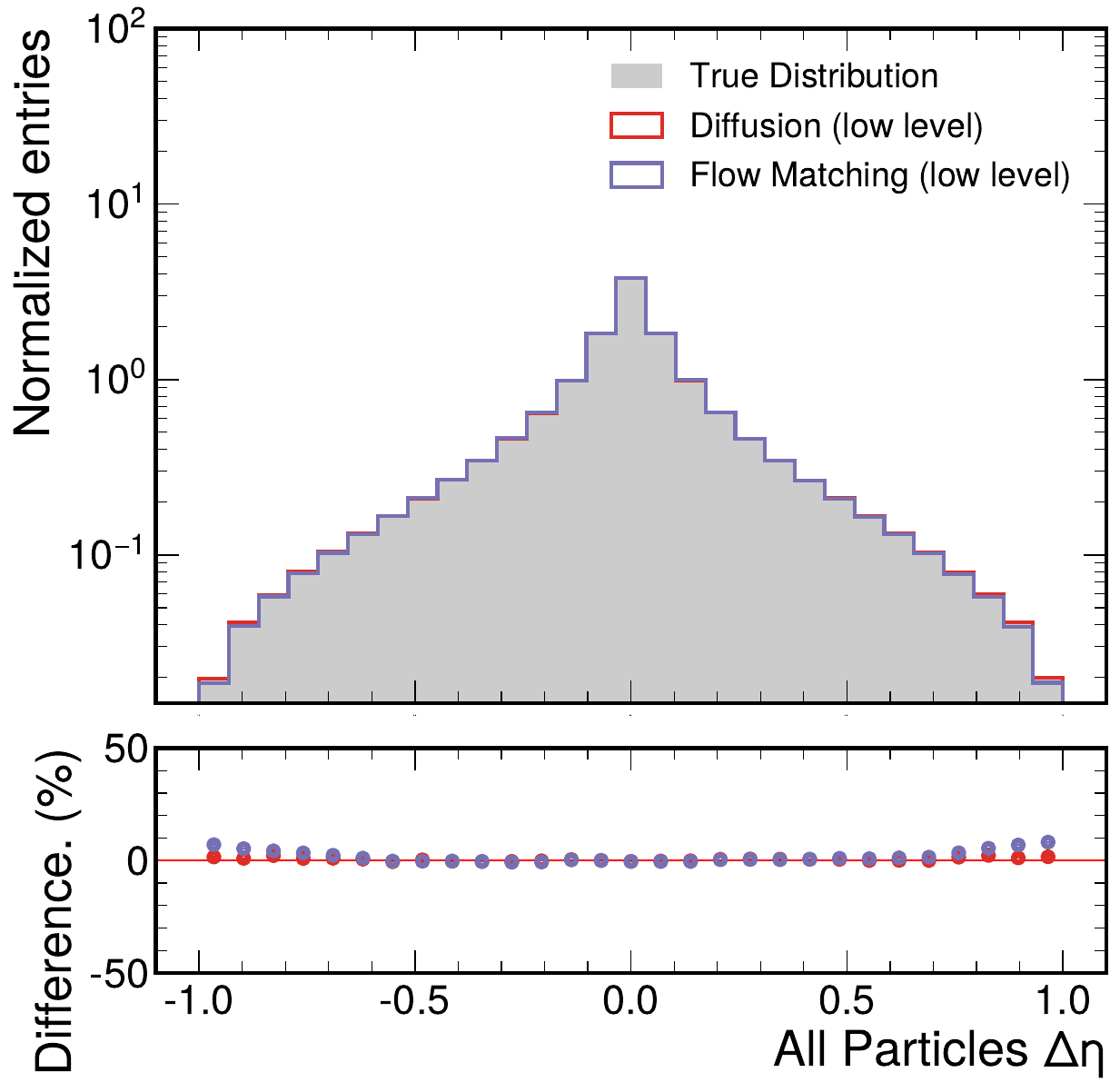}
    \includegraphics[width=0.23\textwidth]{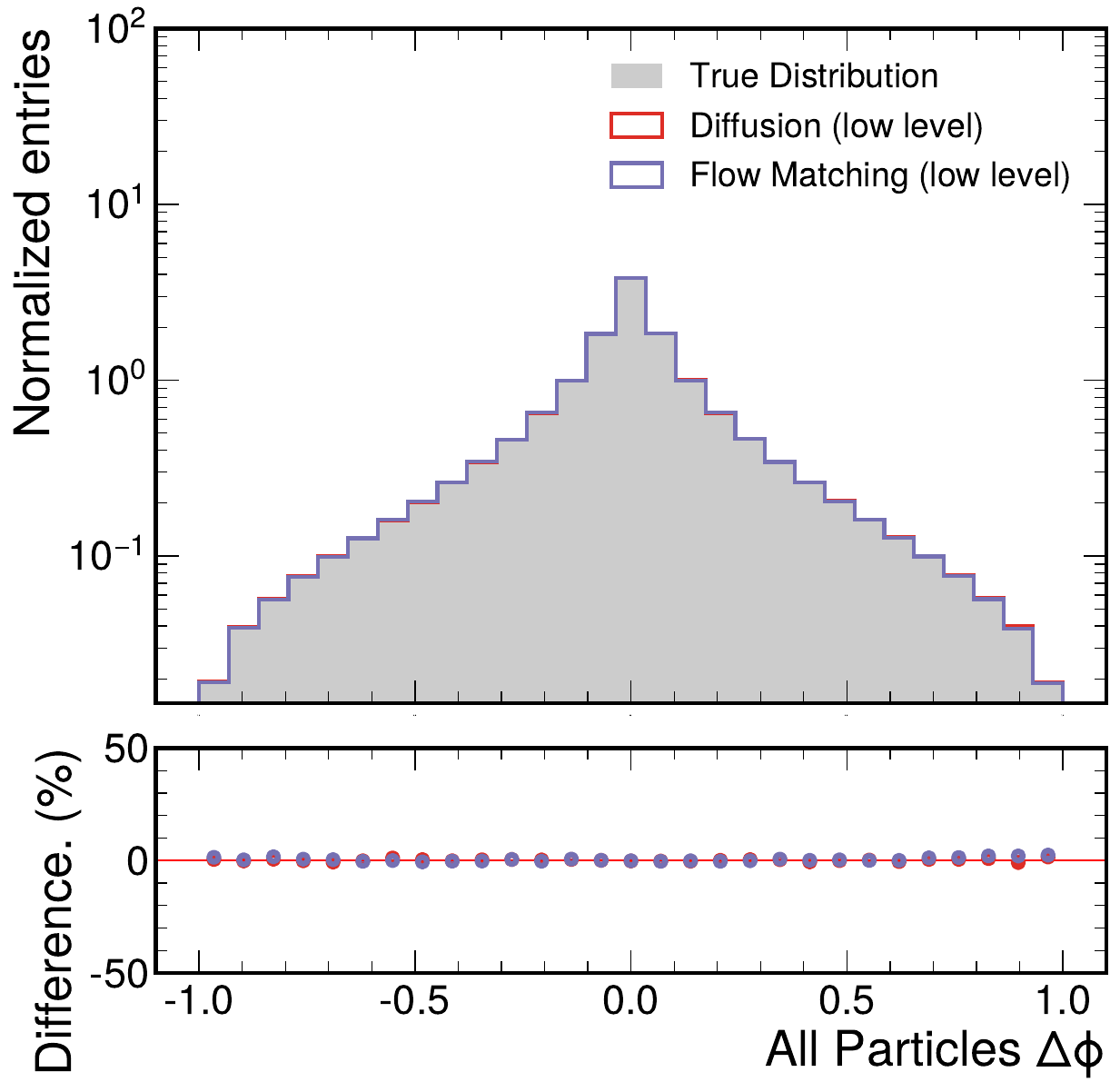}
\caption{Comparison between jet (first five histograms) and particle (last three histograms) kinematic distributions for samples in the dijet mass sideband. Distributions generated by the diffusion and flow matching models are shown in red and blue respectively, while expected distributions from the pseudo-data are shown as filled histograms.}
\label{fig:hist_SB}
\end{figure*}

We evaluate the performance of the generative model first by comparing distributions of generated quantities with the ones present in the simulation of sideband events. Histograms of the jet and particle kinematic information are shown in Fig.~\ref{fig:hist_SB}. We observe a good agreement between generated and simulated quantities, often to better than 10\% accuracy. 

 We further evaluate the quality of the generative models using a binary classifier  trained to 
distinguish between generated and reference samples, as proposed in \cite{Krause:2021ilc}. Using the same classifier architecture described in Sec.~\ref{sec:method}, trained on generated vs.\ reference events in the sideband region, we obtain an area under the curve (AUC) metric calculated over 10 independent trainings of $0.514\,\pm\,0.07$ and $0.522\,\pm\,0.006$ for the diffusion and flow matching models respectively, values very close to the best attainable result of $0.50$ representing the scenario where the classifier is unable to tell the difference between the samples. Uncertainties of the AUC values are derived from 10 independent trainings of the classifier using different random initialization.

\begin{figure*}[ht]
\centering
    \includegraphics[width=0.23\textwidth]{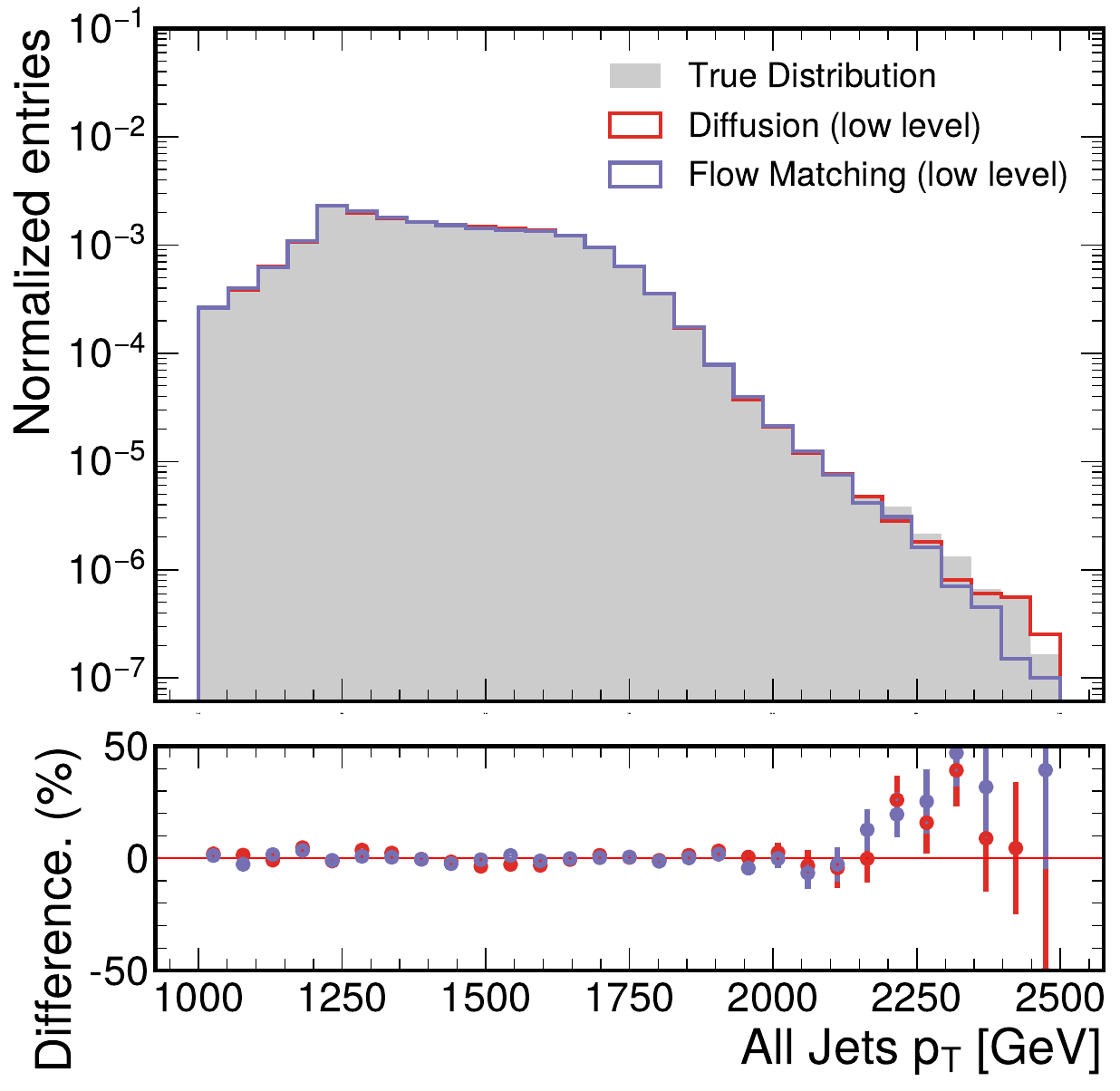}
    \includegraphics[width=0.23\textwidth]{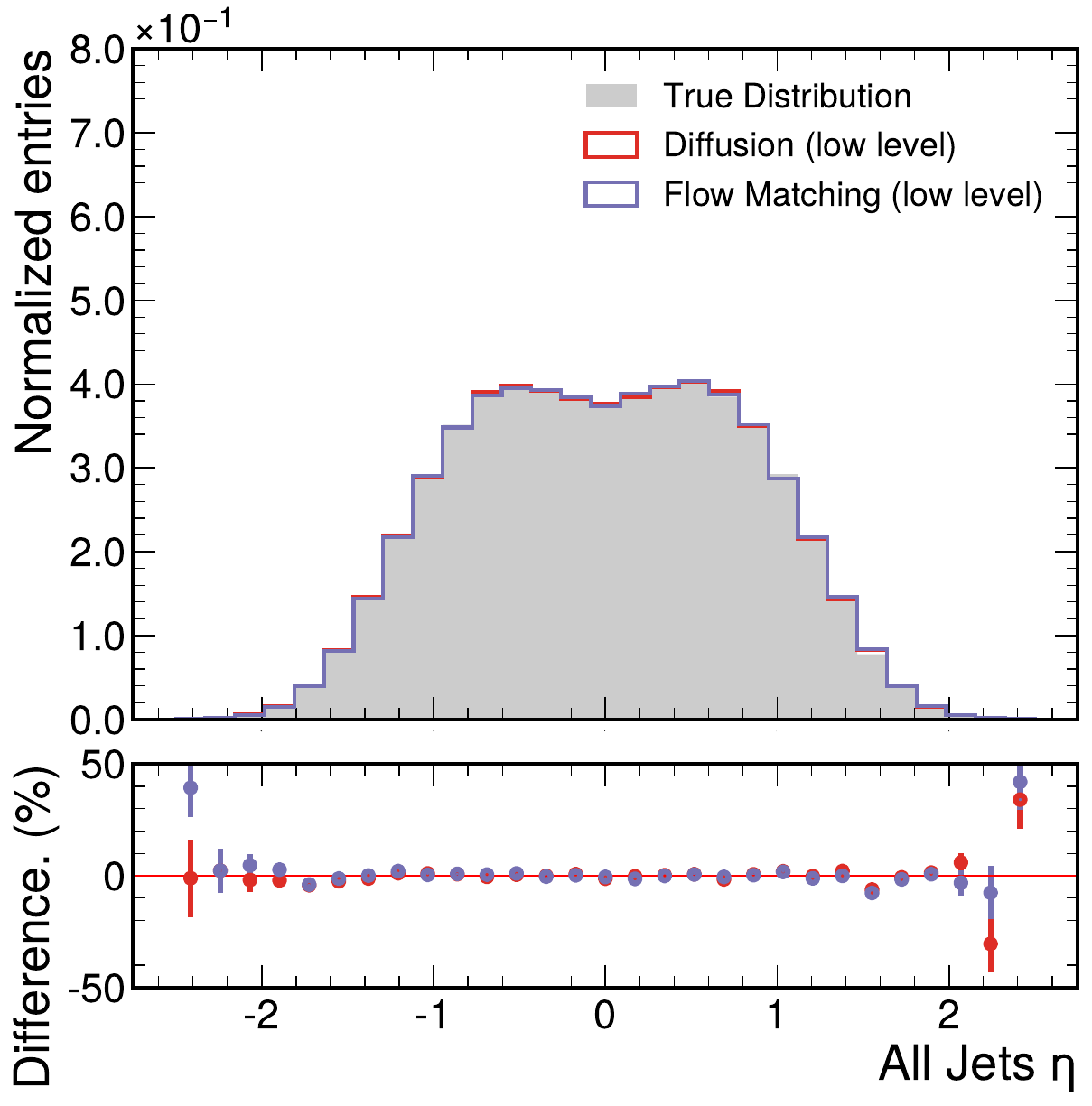}
    \includegraphics[width=0.23\textwidth]{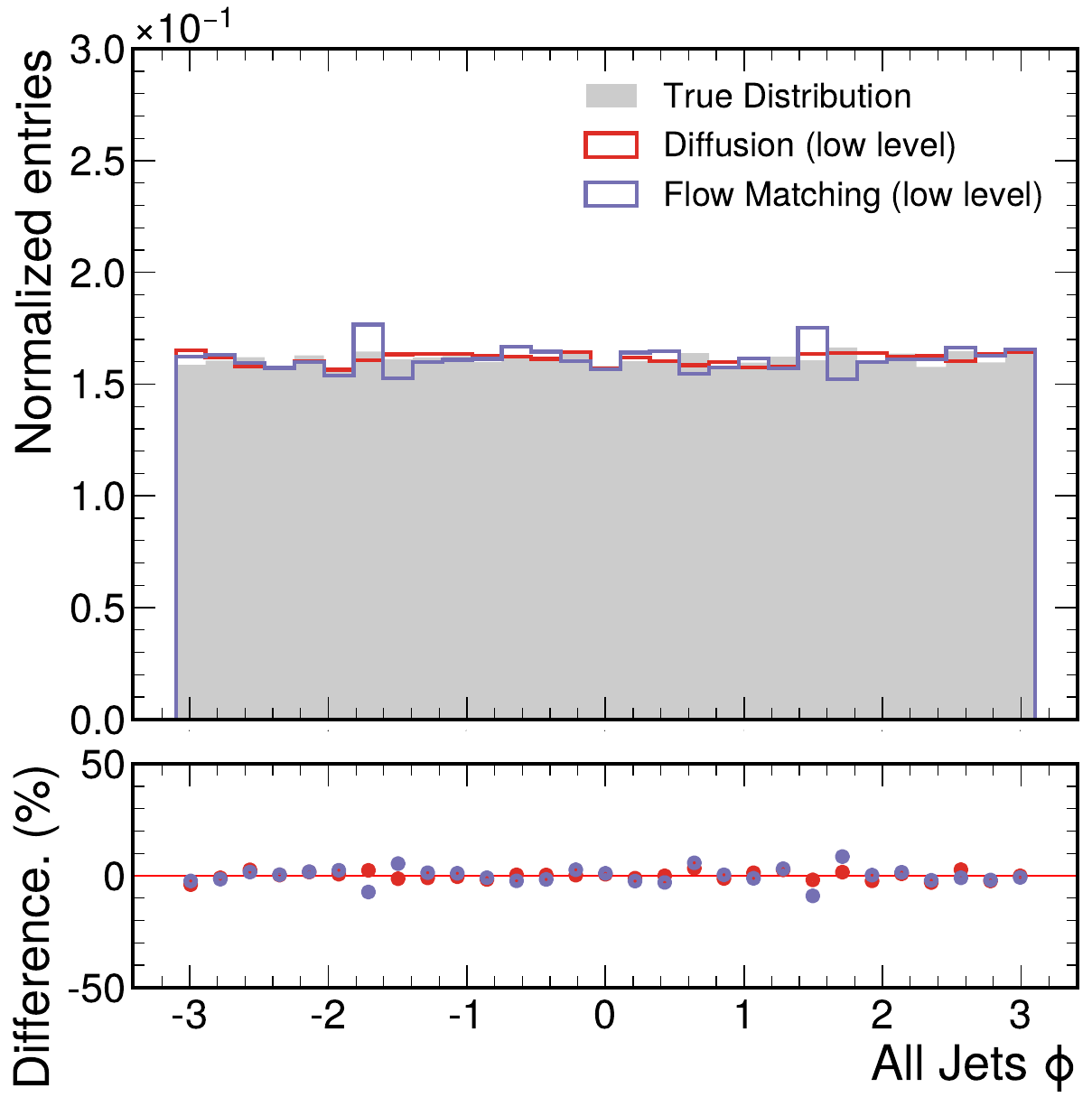}
    \includegraphics[width=0.23\textwidth]{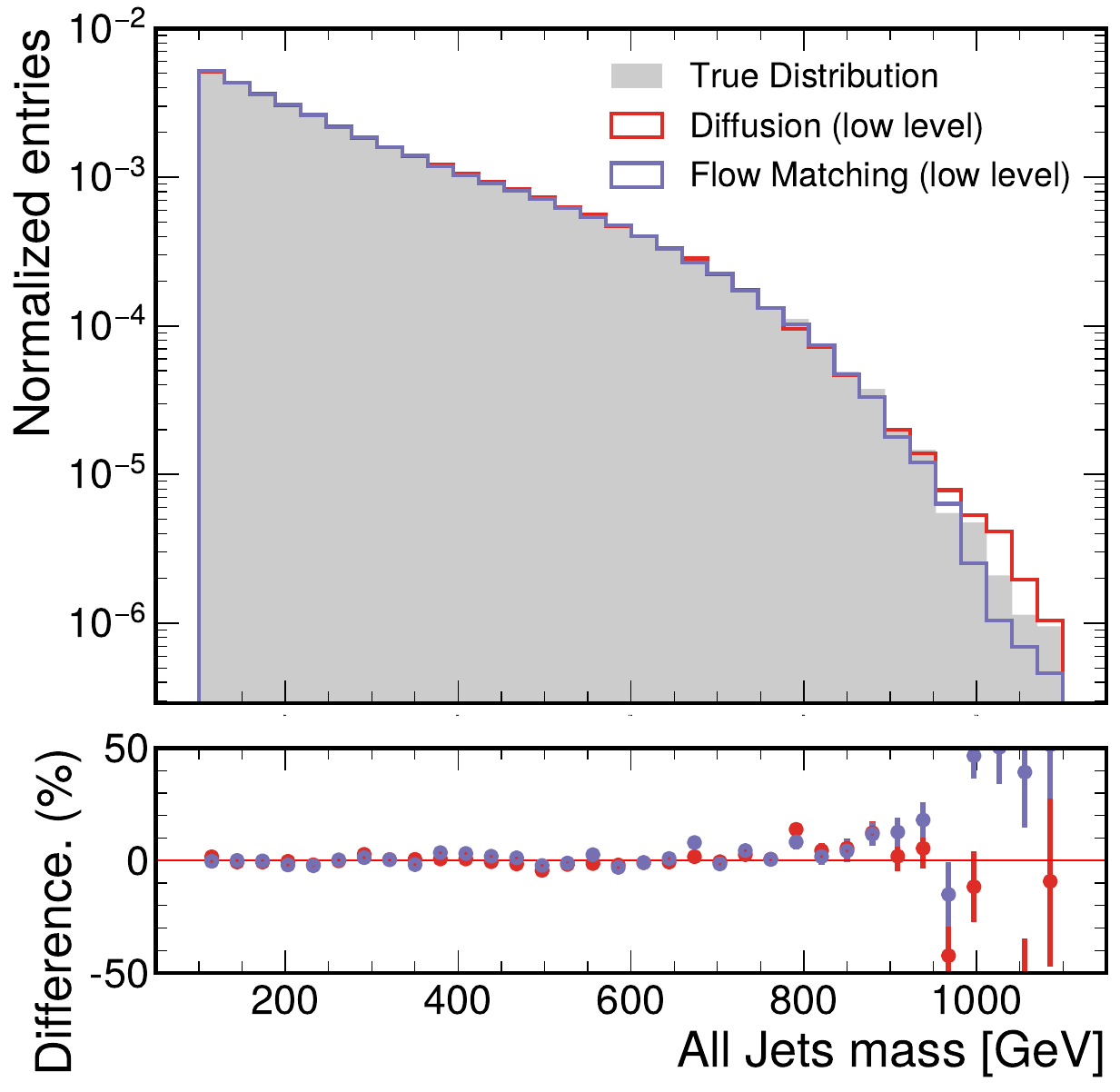}
    \includegraphics[width=0.23\textwidth]{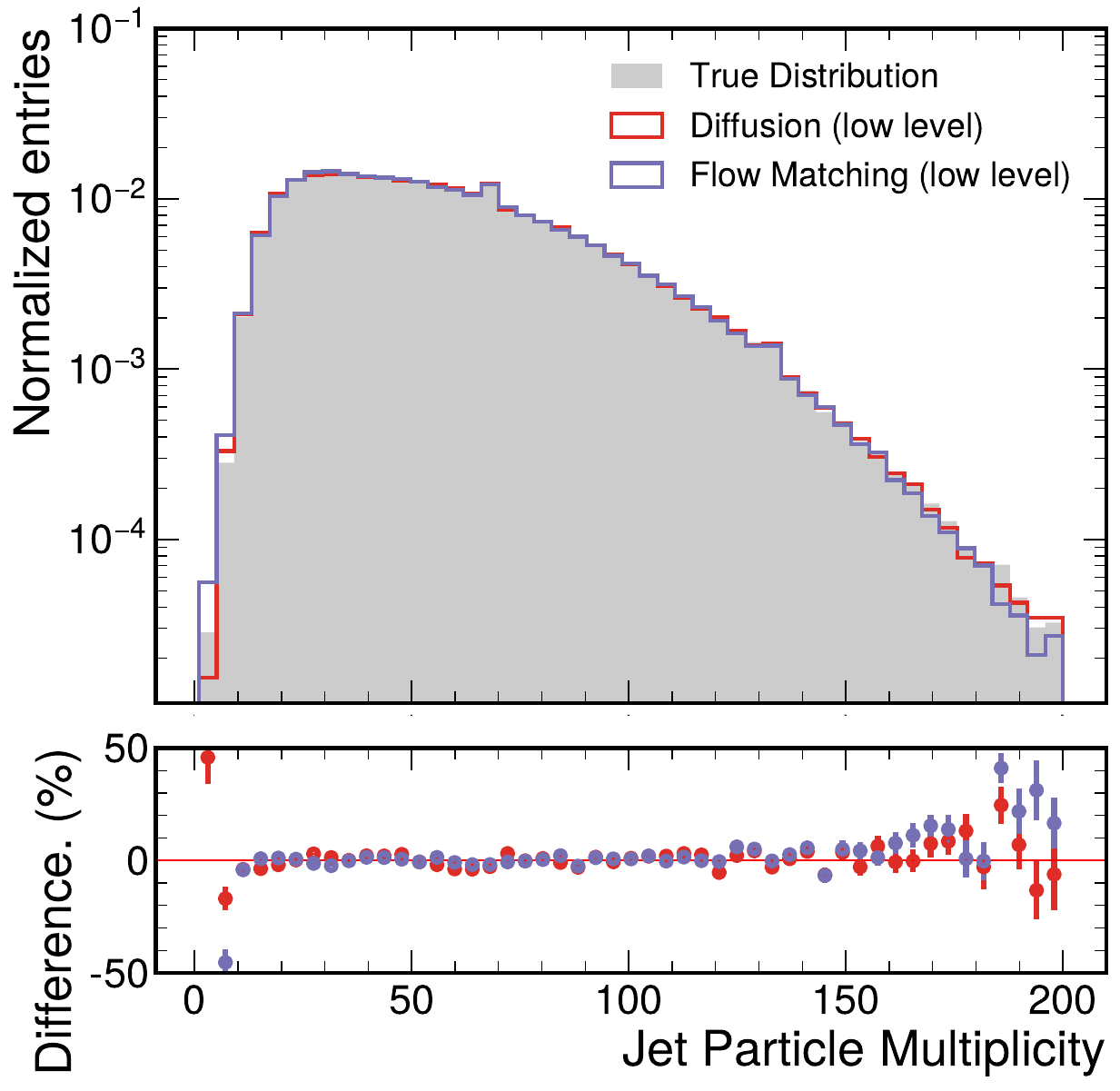}
    \includegraphics[width=0.23\textwidth]{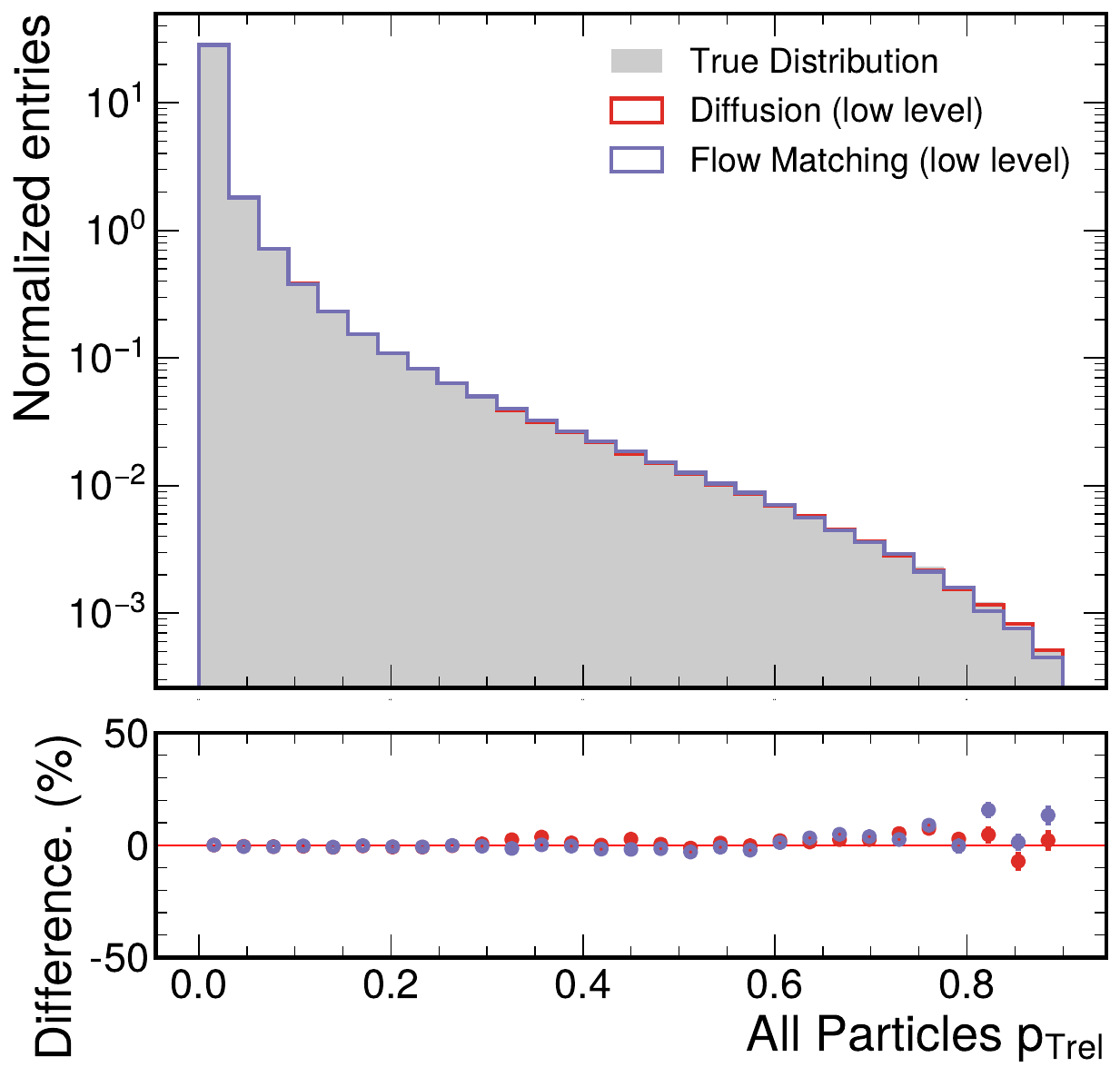}
    \includegraphics[width=0.23\textwidth]{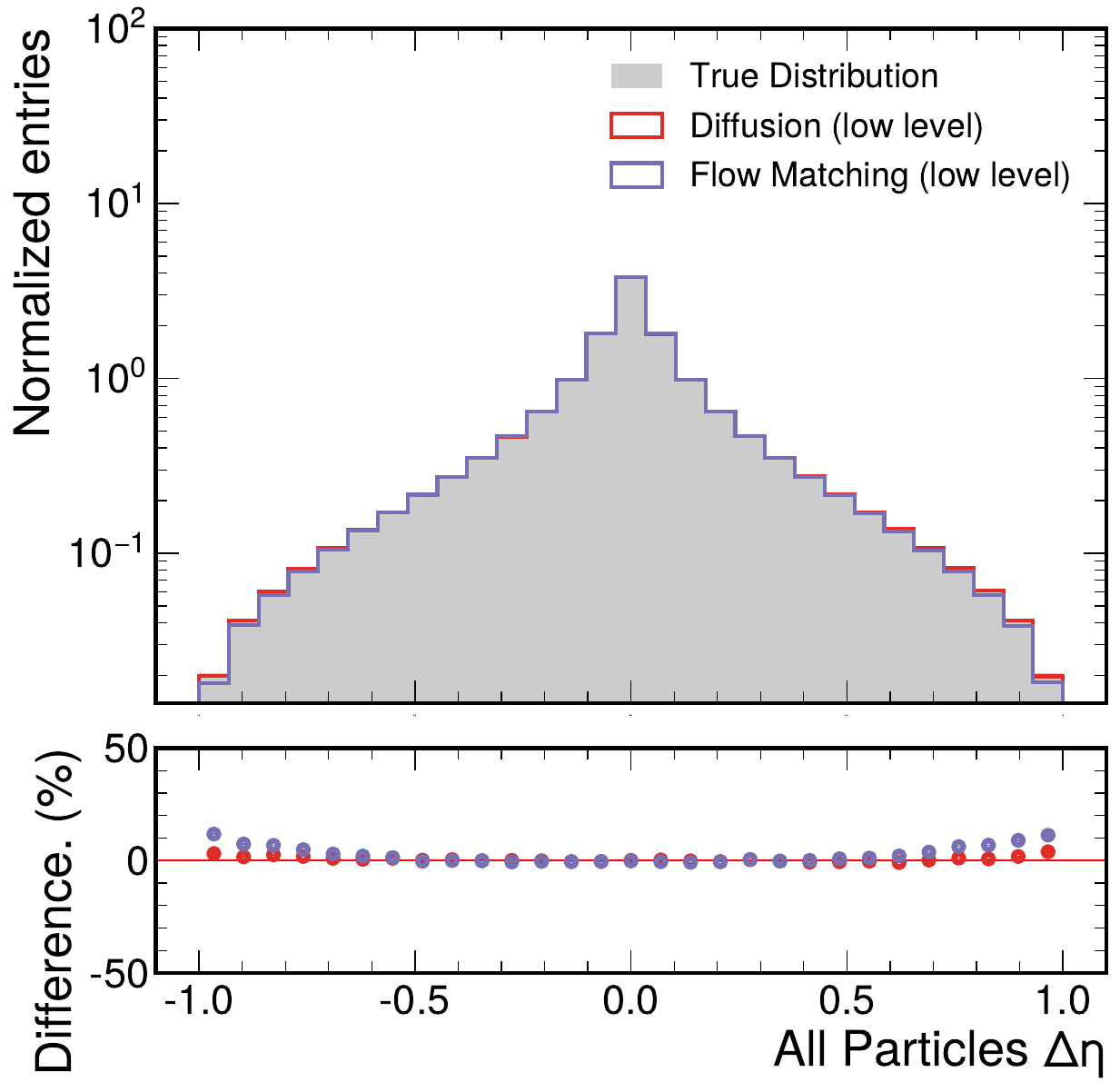}
    \includegraphics[width=0.23\textwidth]{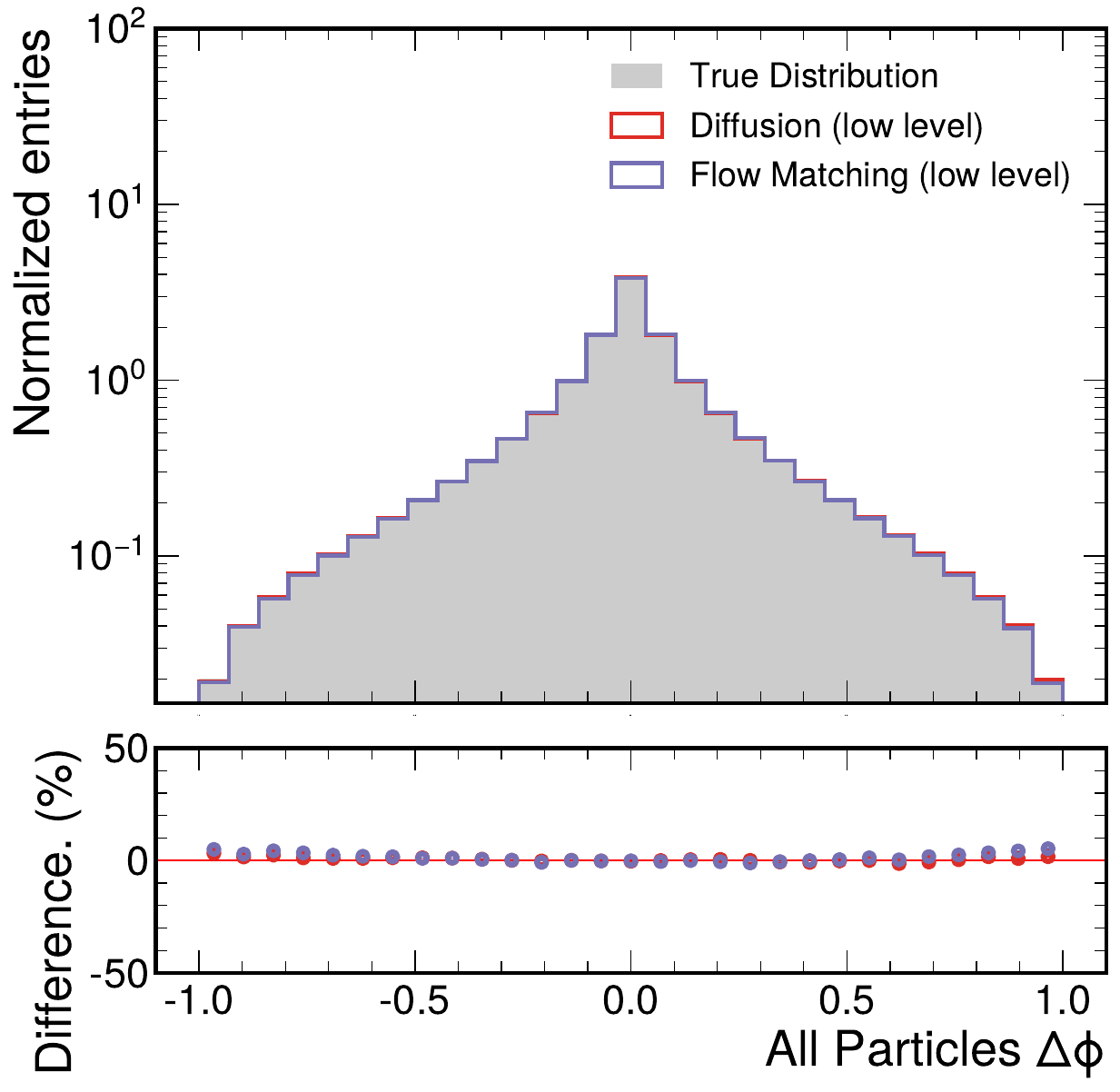}
\caption{Comparison between jet (first five histograms) and particle (last three histograms) kinematic distributions for samples in the dijet mass signal region. Distributions generated by the diffusion and flow matching models are shown in red and blue respectively, while expected background distributions from the pseudo-data are shown as filled histograms.}
\label{fig:hist_SR}
\end{figure*}

The next step is to generate background observations in the signal region. Values of $m_{jj}$ are first generated by sampling values from a kernel density estimator trained to learn the probability density of the $m_{jj}$ spectra based on the values of the observations coming from the sideband regions. The values sampled in the signal region are then used as conditional inputs to the generative models. A total of 200k background events are generated in the signal region, roughly a factor of two more than the amount of true background events in the data within the signal region.\footnote{In \cite{Hallin:2021wme} an even larger amount of oversampling, i.e.\ a larger number of synthetic background events, was found to be helpful in improving the performance of the weakly-supervised classifier. It would be interesting to study that here as well, but we save this for future work.}
A comparison of the simulated and generated background distributions in the signal region is shown in Fig.~\ref{fig:hist_SR}. We observe a good agreement between simulated and generated background events with only noticeable differences observed at the extremes of the relative $\eta$ and $\phi$ distributions. In the absence of signal, the classifier trained to distinguish between generated vs. reference samples yield an AUC of $0.52\,\pm\, 0.01$ and $0.53\,\pm\,0.01$ for the diffusion and flow matching models respectively. 

As a last step, we derive anomaly scores by training the classifier in the signal region to distinguish the neural network-generated background events from the data observed in the signal region.
%with different amounts of signal injection. 
During the training, 200k generated samples are trained against the data consisting of 100k background events and different amount of injected signal. From this dataset, 80\% is used for training and 20\% for validation. The training is stopped when the loss in the validation set does not decrease for 20 consecutive epochs. For each different signal injection value we train ten classifiers with different random initialization. 

The evaluation of the classifier is performed using an independent dataset \cite{LHCOlympics,extraLHCOdataset} consisting of 100k signal and background events, used to study different performance metrics.  From the classifier outputs, we calculate the significance improvement characteristic curve (SIC) defined as the signal efficiency divided by the square root of the background efficiency versus the signal efficiency. The SIC represents a multiplicative scaling by which the significance of a signal would increase corresponding to a threshold cut given by a particular signal efficiency.  

\begin{figure*}[ht]
\centering
    \includegraphics[width=0.48\textwidth]{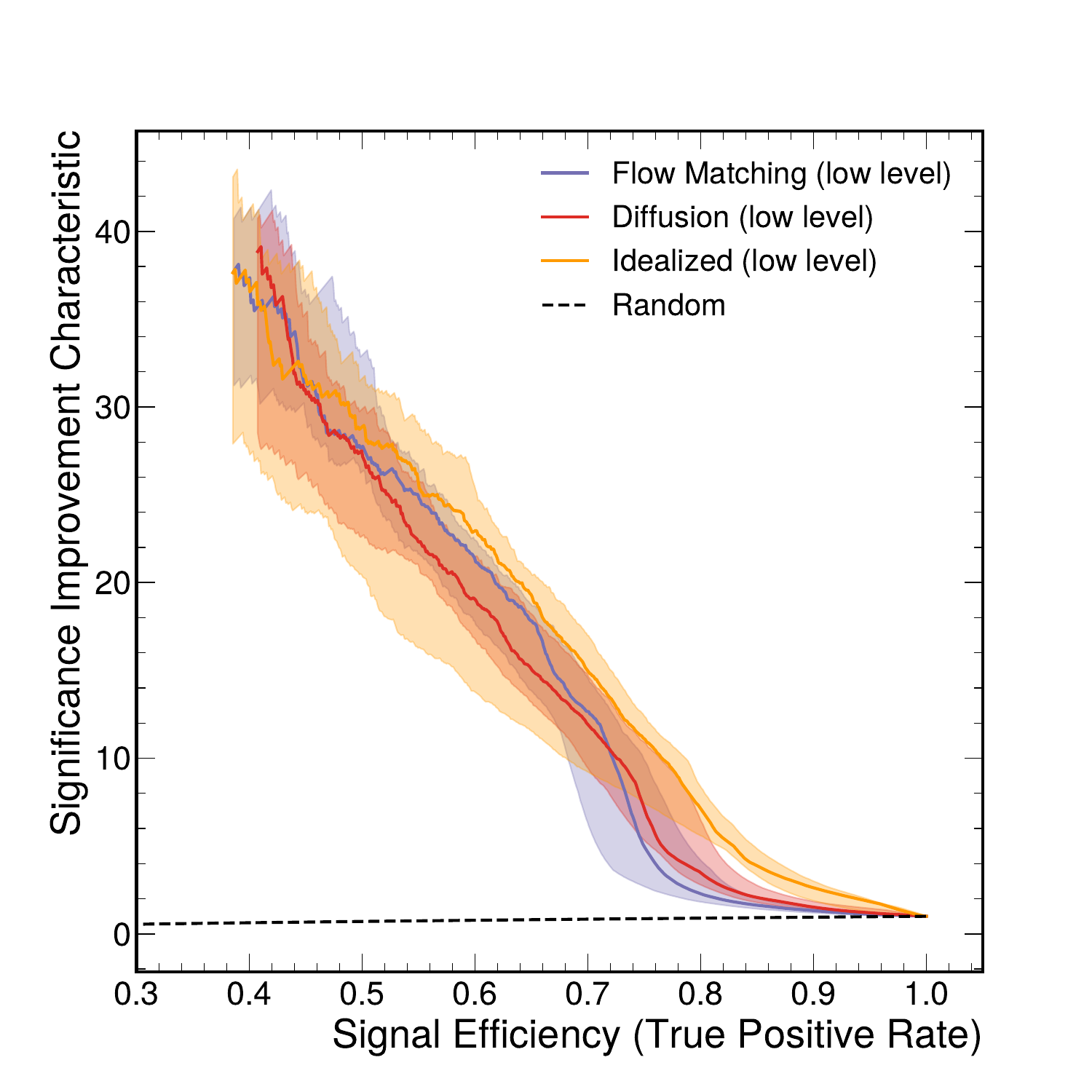}
    \includegraphics[width=0.48\textwidth]{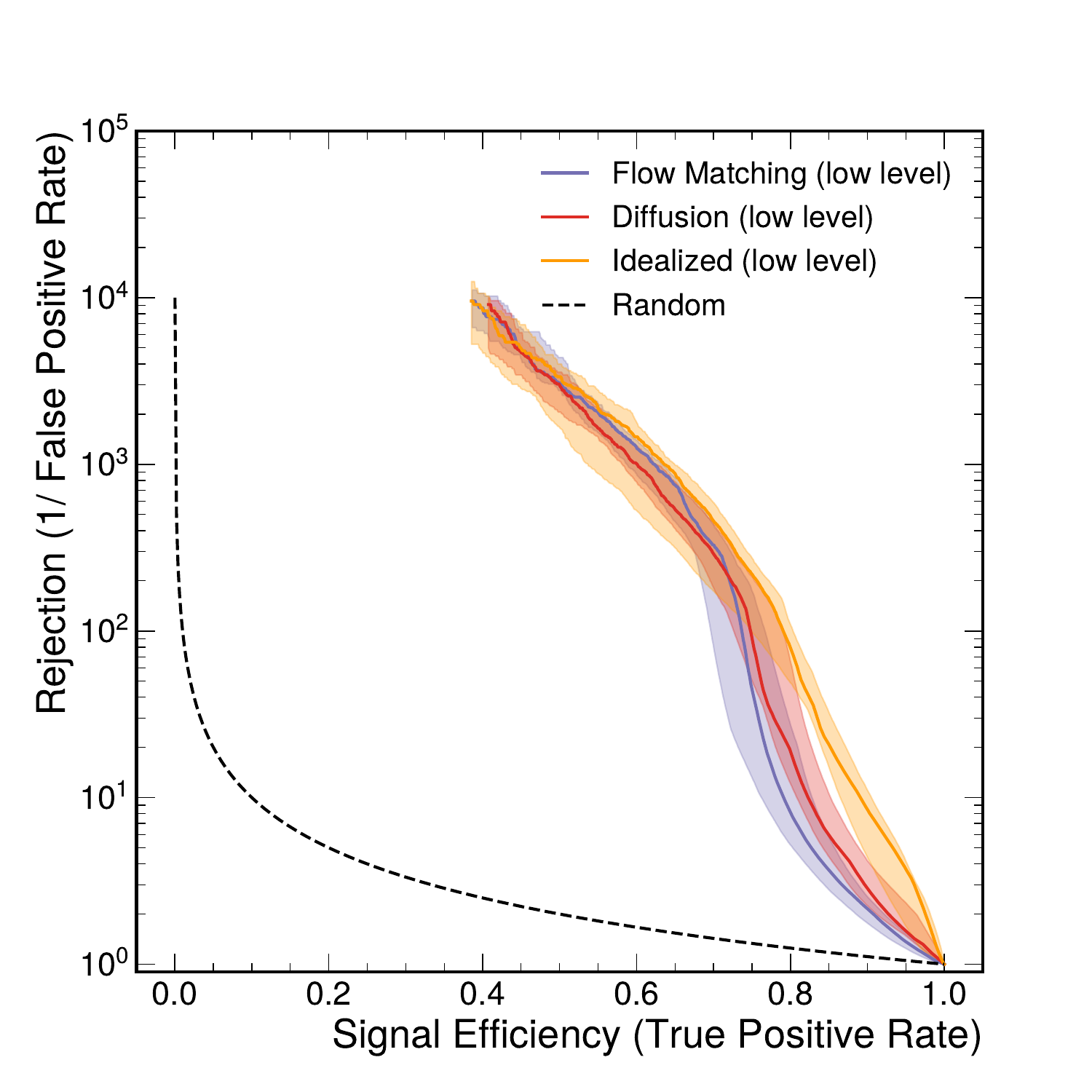}
\caption{Significance improvement characteristic curve (SIC) (left) and  receiver operating characteristic (ROC) (right) obtained by a classifier trained on a sample containing 2000 signal and 100k background events. SIC and ROC values are left empty if the number of background events is below 10  after applying a given threshold with the classifier output to avoid large statistical fluctuations.}
\label{fig:sic_roc}
\end{figure*}

In Fig.~\ref{fig:sic_roc}, we show an example of the median of the receiver operating characteristic (ROC) and SIC curves, obtained for a signal region containing 2000 signal and 100k background events. The uncertainty band is derived from the 68\% confidence bands of 10 independent classifier trainings.  The values are compared with an idealized anomaly detector \cite{Hallin:2021wme} created by training the classifier with independent background events simulated with the same physics simulator used to generate the background in the signal region. A total of 200k background events are used to train the classifier taken from an independent background sample~\cite{extraLHCOdataset}. The idealized classifier training follows the same training strategy discussed before.   The evaluation is always performed on a large set of signal events and independent background sample~\cite{extraLHCOdataset}, to mitigate the statistical fluctuations from the signal injection. Both SIC and AUC values are always calculated using the signal and background efficiencies.

\begin{figure*}[ht]
\centering
    \includegraphics[width=0.48\textwidth]{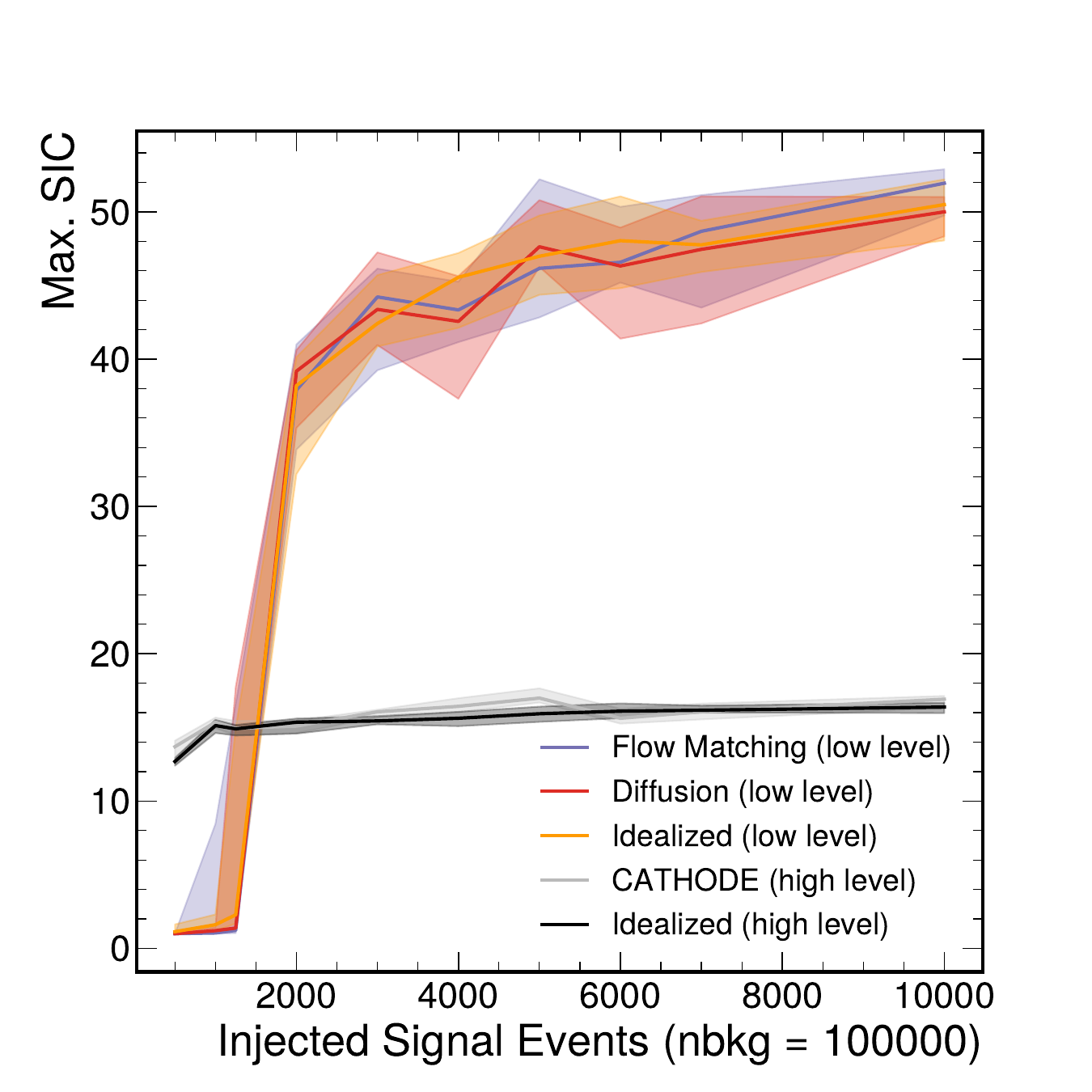}
    \includegraphics[width=0.48\textwidth]{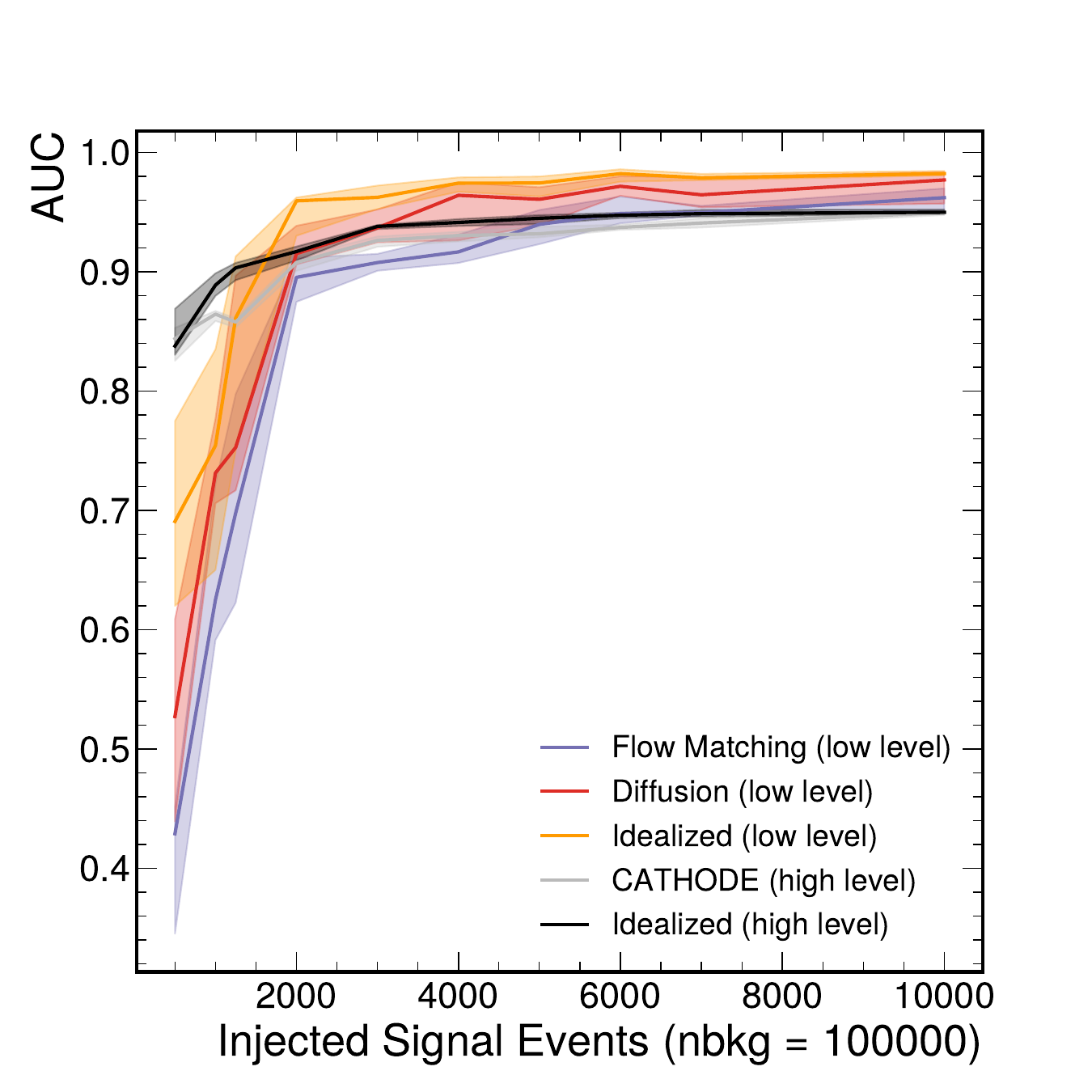}
\caption{Median of the Maximum value of the significance improvement characteristic (SIC) curve (left) and area under the curve of the receiver operating characteristic (AUC) (right) as a function of the number of injected signal events. Uncertainties are derived from the 68\% confidence bands calculated over ten independent classifier trainings with random initialization. }
\label{fig:max_sic_auc}
\end{figure*}

Both generative models show excellent performance at identifying the signal component, showing a maximum value of the SIC curve near 40, limited by the amount of background events available in the signal region. We also investigate the maximum SIC value and AUC obtained by training the classifier with different amounts of signal injection. The results are shown in Fig.~\ref{fig:max_sic_auc}, compared with the idealized anomaly detector results and with   a retrained version of the original \textsc{Cathode} method~\cite{Hallin:2021wme} in Fig.~\ref{fig:max_sic_auc}.

%the idealized anomaly detector results and with 

At lower amounts of signal injection, low-dimensional \textsc{Cathode} outperforms even the idealized anomaly detector due to the high dimensionality of the inputs to the latter vs.\ the hand-picked high-level features input to the former.

Above 1k signal injected events, the idealized anomaly detector matches the maximum performance obtained by the low-dimensional \textsc{Cathode} and continues to improve as the amount of signal increases. Similar conclusions are made for the diffusion and flow matching generative models, which are able to reach a maximum SIC value as high as 50 for large amounts of signal injection. At lower amounts of signal injection, the AUC of the ROC curve obtained by the models is compatible with 50\%, showing that the background prediction in the signal region is precise.

\section{Conclusions and Outlook}
\label{sec:conclusions}

Anomaly detection provides a new set of tools for exploring particle physics data in search of BSM physics.  Machine learning is empowering these approaches to make the most of our complex datasets.  We have explored how state-of-the-art generative neural networks can be deployed in a resonant anomaly detection mode to use all of the available information in hadronic final states. On the LHC Olympics dataset, we find that this method is able to automatically identify the presence of anomalies, even in the high- and variable-dimensional feature space so long as there is enough signal.  This performance comes at the tradeoff of specificity.  A simpler model trained on a small set of features known to be useful for the signal performs better at low signal injection and worse at relatively high signal injection.  This crossover point currently happens around 5-6$\sigma$ in injected signal, but future innovations may be able to bring this down.  It will also be exciting to explore the breadth of signals that this new approach can find.  Quantifying the discovery volume  will be a complex challenge that is already difficult for existing methods and ever more acute with the full phase space.

\section*{Code Availability}
The code used to produce all results presented in this paper are available at \url{https://github.com/ViniciusMikuni/LHCO_diffusion} and \url{https://github.com/uhh-pd-ml/LHCO_EPiC-FM}.

\section*{Acknowledgments}

The authors would like to thank Manuel Sommerhalder for providing reference results for low-dimensional \textsc{Cathode}; Tobias Quadfasel for assistance in preparing some figures; and Darius Faroughy, Julie Pagès and Manuel Szewc for early discussions.
VM and BN are supported by the U.S. Department of Energy (DOE), Office of Science under contract DE-AC02-05CH11231. DS is supported by DOE grant DOE-SC0010008. This research used resources of the National Energy Research Scientific Computing Center, a DOE Office of Science User Facility supported by the Office of Science of the U.S. Department of Energy under Contract No. DE-AC02-05CH11231 using NERSC award HEP-ERCAP0021099.  
EB is funded by a scholarship of the Friedrich Naumann Foundation for Freedom and by the German Federal Ministry of Science and Research (BMBF) via Verbundprojekts 05H2018 - R\&D COMPUTING (Pilotmaßnahme ErUM-Data) Innovative Digitale Technologien für die Erforschung von Universum und Materie.
EB, CE, and GK acknowledge support by the Deutsche Forschungsgemeinschaft under Germany’s Excellence Strategy – EXC 2121  Quantum Universe – 390833306 
and via the KISS consortium (05D23GU4) funded by the German Federal Ministry of Education and Research BMBF in the ErUM-Data action plan.
This research was supported in part through the Maxwell computational resources operated at Deutsches Elektronen-Synchrotron DESY, Hamburg, Germany.

\bibliography{HEPML}
\bibliographystyle{apsrev4-1}

\clearpage
\appendix

\begin{table*}[htbp]
\centering
\caption{Hyperparameter choices for all models. The conditioning variables refer to time $t$, jet variables $p_T$, $\eta$, $\phi$, $m$, particle multiplicity $N$, dijet mass $m_{jj}$.}
\label{tab:hyperparameters}
\resizebox{\textwidth}{!}{
\begin{tabular}{l c c c c c}
\hline
 Hyperparameter & Jet-Diffusion & Particle-Diffusion & Jet-FM & Particle-FM & Classifier  \\
 \hline
 \hline
    Explicit Conditioning & $t$, $m_{jj}$ & $t$, $p_T$, $\eta$, $\phi$, $m$, $N$, $m_{jj}$  & $t$, $m_{jj}$ & $t$, $p_T$, $\eta$, $\phi$, $m$, & / \\
    Time Embedding & Fourier~\cite{tancik2020fourier} & Fourier~\cite{tancik2020fourier} & Sin/Cos & Cosine~\cite{Leigh:2023toe} & / \\
    Activation function & LeakyReLU($0.01$)~\cite{DBLP:journals/corr/XuWCL15} & LeakyReLU($0.01$)~\cite{DBLP:journals/corr/XuWCL15} & ELU~\cite{elu} & LeakyReLU($0.01$)~\cite{DBLP:journals/corr/XuWCL15} & LeakyReLU($0.01$)~\cite{DBLP:journals/corr/XuWCL15} \\
    Batch size & 128 & 128 & 128 & 1024 & 128  \\
    Optimizer & Adam~\cite{adam} & Adam~\cite{adam} & AdamW~\cite{adamw} & AdamW~\cite{adamw} & Adam~\cite{adam} \\
    Initial learning rate & $1.6\times 10{-3}$ & $1.6\times10^{-3}$ & $10^{-3}$ & $10^{-3}$ & $10^{-3}$  \\
    Weight decay & / & / & $5\cdot10^{-5}$ & $5\cdot10^{-5}$ & / \\
    Learning rate scheduling & Cosine annealing~\cite{DBLP:journals/corr/LoshchilovH16a} & Cosine annealing~\cite{DBLP:journals/corr/LoshchilovH16a} & Constant &  Cosine with warm-up & /\\
    Warm-up epochs & / & / & / & 500 & / \\
    Training epochs & 300 & 300 & 10000 & 5000 & 300 \\
    Early stopping patience & 50 & 50 & 100 & / & 50  \\
    Number of GPUs & 16 & 16 & 1 & 1 & 16 \\
\hline
    Model weights & $\sim 1.3M $ & $\sim 1.4M$ & $\sim 380k$ & $\sim 2M$ & $438k$ \\
\hline
\end{tabular}
}
\end{table*}
\end{document}